\documentclass[prc,aps,float,twocolumn,superscriptaddress,showkeys,showpacs,nofootinbib]{revtex4}
\pdfoutput=1
\usepackage{amsmath}
\usepackage{amssymb}
\usepackage{amsfonts}
\usepackage{epsfig}
\usepackage{graphics}
\usepackage{color}

\renewcommand{\vec}[1]{\mathbf{#1}}

\usepackage{bbm}
\newcommand\bbone{\ensuremath{\mathbbm{1}}}

\begin{document}

\title{An improved density matrix expansion for spin-unsaturated nuclei}

\author{B. Gebremariam}
\email{gebremar@nscl.msu.edu} \affiliation{National Superconducting
Cyclotron Laboratory, 1 Cyclotron Laboratory, East-Lansing, MI
48824, USA} \affiliation{Department of Physics and Astronomy,
Michigan State University, East Lansing, MI 48824, USA}

\author{T. Duguet}
\email{thomas.duguet@cea.fr} \affiliation{National Superconducting
Cyclotron Laboratory, 1 Cyclotron Laboratory, East-Lansing, MI
48824, USA} \affiliation{Department of Physics and Astronomy,
Michigan State University, East Lansing, MI 48824, USA}
\affiliation{CEA, Centre de Saclay, IRFU/Service de Physique
Nucl{\'e}aire, F-91191 Gif-sur-Yvette, France}

\author{S. K. Bogner}
\email{bogner@nscl.msu.edu} \affiliation{National Superconducting
Cyclotron Laboratory, 1 Cyclotron Laboratory, East-Lansing, MI
48824, USA} \affiliation{Department of Physics and Astronomy,
Michigan State University, East Lansing, MI 48824, USA}

\date{\today}

\pacs{21.10.Re, 21.60.Ev, 71.15.Mb}

\keywords{Density matrix expansion, non-empirical energy density functional}

\begin{abstract}
A current objective of low-energy nuclear theory is to build non-empirical nuclear energy density functionals (EDFs) from underlying inter-nucleon interactions and many-body perturbation theory (MBPT). The density matrix expansion (DME) of Negele and Vautherin is a convenient method to map highly non-local Hartree-Fock expressions into the form of a quasi-local Skyrme functional with density-dependent couplings. In this work, we assess the accuracy of the DME at reproducing the non-local exchange (Fock) contribution to the energy. In contrast to the scalar part of the density matrix for which the original formulation of Negele and Vautherin is reasonably accurate, we demonstrate the necessity to reformulate the DME for the vector part of the density matrix, which is needed for an accurate description of spin-unsaturated nuclei. Phase-space averaging techniques are shown to yield a significant improvement for the vector part of the density matrix compared to the original formulation of Negele and Vautherin. The key to the improved accuracy is to take into account the anisotropy that characterizes the local-momentum distribution in the surface region of finite Fermi systems. Optimizing separately the DME for the central, tensor and spin-orbit contributions to the Fock energy, one reaches a few-percent accuracy over a representative set of semi-magic nuclei. With such an accuracy at hand, one can envision using the corresponding Skyrme-like energy functional as a microscopically-constrained starting point around which future phenomenological parameterizations can be built and refined.



\end{abstract}

\maketitle

%
%

\section{Introduction}
\label{intro}

The nuclear energy density functional (EDF) approach is the many-body method of choice to study medium-mass and heavy nuclei in a systematic manner~\cite{bender03b}. Modern parameterizations of empirical energy functionals (e.g. Skyrme, Gogny or their relativistic counterparts) provide a fair description of bulk properties and certain spectroscopic features of known nuclei. However, such empirical EDFs lack predictive power and a true spectroscopic quality away from known data. Consequently, an intense ongoing effort is dedicated to empirically improving the analytical form and the fitting of energy density functionals~\cite{lesinski06a,Lesinski:2007zz,Margueron:2007uf,Niksic:2008vp,carlsson09,Goriely:2009zz}.

A complementary approach in the quest for predictive EDFs~\cite{Lesinski:2008cd,Drut:2009ce,Duguet:2009gc,Bogner:2008kj,Kaiser:2003uh} relies less on fitting empirical functionals to known data, but rather attempts to constrain the analytical form of the functional and the values of its couplings from many-body perturbation theory (MBPT) and the underlying two- and three-nucleon (NN and NNN) interactions. Switching from conventional hard-core potentials to low-momentum interactions is essential in this respect, as the many-body problem formulated in terms of the latter becomes significantly more perturbative\footnote{The need for infinite resummation of certain sets of diagrams and/or the redefinition of the unperturbed vacuum $|\Phi\rangle$ cannot be ruled out at this point.}. Indeed, second-order perturbative calculations provide a good account of bulk correlations in both infinite nuclear matter~\cite{Bogner:2009un} and doubly-magic nuclei~\cite{Roth:2005ah}. Using many-body perturbation theory (MBPT)~\cite{blaizot86} as a baseline, the long term goals of the project are to (i) bridge non-empirical EDF methods with {\it ab-initio} many-body techniques applicable to light nuclei, (ii) calculate properties of heavy/complex nuclei from basic vacuum interactions and (iii) perform EDF calculations with controllable theoretical errors.

MBPT contributions to the energy are written in terms of density matrices and propagators convolved with finite-range interaction vertices, and are therefore highly non-local in both space and time. In order to make such functionals numerically tractable in heavy open-shell nuclei, it is desirable to develop simplified approximations expressed in terms of the local densities and currents. Starting at lowest order, which displays only non-locality in space through the Fock contribution to the energy\footnote{For simplicity, we are assuming local NN and NNN interactions.} , the objective of the present work is to revisit the density matrix expansion (DME) of Negele and Vautherin~\cite{negele72} to assess its accuracy in reproducing non-local Fock contributions.


The focus of the present paper is on the vector part of the density matrix, which is relevant for approximating the central, tensor and spin-orbit Fock contributions in spin-unsaturated nuclei, i.e. in nuclei where only one of two spin-orbit partners is filled. Indeed, the few tests of the DME over the past thirty-five years have focused entirely on the scalar part~\cite{sprung75,hofmann97,Bogner:2008kj}, given that no reliable expansion of the vector part of the density matrix was ever proposed. As acknowledged by Negele and Vautherin in their seminal paper, the expansion suggested for the vector part of the density matrix was not on the same level as the one designed for its scalar part. Such a feature is obviously critical since the overwhelming majority of nuclei are spin unsaturated. Here, we demonstrate that phase-space averaging techniques allow a consistent expansion of both the scalar and the vector parts of the density matrix, such that the accuracy is greatly improved for the latter. A key feature of the new method is to take into account the deformation displayed by the local momentum distribution at the surface of most finite fermi systems~\cite{durand82,bulgac96}. While it is shown to have little impact on the expansion of the scalar part, the deformation of the local momentum distribution is crucial to accurately reproduce contributions to the energy that probe the vector part of the density matrix.


The paper is organized as follows. Section~\ref{inputspaper} provides the basic ingredients needed to conduct the present study. Section~\ref{PSA-refomulation} is dedicated to the reformulation of the density matrix expansion on the basis of phase-space averaging techniques. The accuracy of the approximation method is gauged in Sec.~\ref{section:results} through non-self consistent tests that make use of two schematic nucleon-nucleon interactions and of density matrices obtained from self-consistent EDF calculations of a large set of semi-magic nuclei. Each of the central, tensor and spin-orbit contributions to the Fock energy is analyzed separately. Conclusions are given in Sec.~\ref{conclusions} while appendices provide complete sets of formulae and analytical derivations. In particular, couplings of the generalized Skyrme-like EDF obtained through the DME (see Eq.~\ref{eqn:skryme-like-EDF}) are provided in appendix~\ref{appendix:explicit-coupling-forms}.

\begin{table}
\begin{center}
\begin{tabular}{ll}
\hline \hline \\
EDF &   Energy density functional \\
DME &   Density matrix expansion\\
PSA &   Phase space averaging\\
OBDM &  One-body density matrix\\
INM &   Infinite nuclear matter\\
& \\
\hline \hline
\end{tabular}
\caption{List of acronyms repeatedly used in the text.}
\label{table:acronym}
\end{center}
\end{table}

%
%
\section{Density matrix and HF energy}
\label{inputspaper}

Let us consider a product state of reference $| \Phi\rangle$. As briefly explained in Sec.~\ref{section:strategy}, this typically is the unperturbed many-body state around which perturbation theory is performed or, in a more phenomenological language, the auxiliary state in terms of which one builds a so-called single-reference energy density functional (EDF). In the present case, we consider an implementation without explicit treatment of superfluidity such that $| \Phi\rangle$ takes the form of a Slater determinant. In addition, we consider the system to be invariant under time-reversal.

\subsection{The one-body density matrix}
\label{illdefined}

The one-body density matrix (OBDM) $\rho$ of the many-body state $| \Phi\rangle$ is defined in terms of operators ${c}^{\dagger} \, (\vec{r} \, \sigma \, q)/{c} \, (\vec{r} \, \sigma \, q)$ that create/annihilate a nucleon at a given position in space $\vec{r}$ with given spin and isospin projections $\sigma=\pm 1/2$ and $q=n,p$ on the quantization axis
\begin{eqnarray}\label{one-body-density-matrix}
\rho_{q} (\vec{r} \, \sigma, \vec{r} ' \, \sigma ')  &\equiv& \langle \Phi | \, {c}^{\dagger} (\vec{r} ' \, \sigma '
\, q) \, {c} \, (\vec{r} \, \sigma \, q) \, | \Phi
\rangle \nonumber\\
& =& \sum_{ij} \, \varphi_{i}^{\ast} (\vec{r}  ' \sigma ' q) \,
\, \varphi_{j} (\vec{r} \sigma q) \, \rho^{q}_{ji} \, \, \, ,
\end{eqnarray}
where it is assumed that single-particle states do not mix isospin projections so that the OBDM is diagonal in isospin space\footnote{The Slater determinant can however break spatial symmetries.}. In Eq.~\ref{one-body-density-matrix},  $\rho^{q}_{ji} \equiv \langle \, \Phi \, | \,
{c}^{\dagger}_{i} \, {c}_{j} \, | \, \Phi \rangle$ defines the
OBDM in an alternate single-particle basis $\{{c}_{i}; \varphi_{i}(\vec{r} \sigma q)\}$. Choosing the particular basis from which $| \Phi\rangle$ is built, $\rho^{q}_{ji}$ becomes diagonal with matrix elements equal to one for occupied states and zero for empty states. The
OBDM can be further separated into
\begin{eqnarray}\label{scavecdef}
\rho_{q} (\vec{r} \, \sigma , \vec{r} ' \, \sigma ')&=&
\frac{1}{2}  \bigl\{\rho_{q} (\vec{r}, \vec{r} ') \, \delta_{\sigma
\sigma '} + \vec{s}_{q} (\vec{r}, \vec{r} ') \cdot \vec{\sigma}_{\sigma \sigma '} \bigr\}\,,
\end{eqnarray}
where the scalar and vector parts are respectively defined as
\begin{eqnarray}
\rho_{q} (\vec{r},\vec{r} ') &\equiv&  \sum_{\sigma \sigma '}
\rho_{q} (\vec{r} \, \sigma , \vec{r} ' \, \sigma^\prime)  \,
\langle \sigma' \lvert  \bbone \rvert \sigma \rangle
 \nonumber\\
&=& \sum_{\sigma}\,\sum_{ij} \, \varphi_{i}^{\ast} (\vec{r} '
\sigma  q) \,
\varphi_{j} (\vec{r} \sigma q)  \,\rho^{q}_{ji}\,\,\,, \label{scalarisoscalarden}\\
\vec{s}_{q} (\vec{r},\vec{r} ') &\equiv& \sum_{\sigma \sigma '}
\rho_{q} (\vec{r} \, \sigma , \vec{r} ' \, \sigma ') \,
\langle \sigma' \lvert \vec{\sigma} \rvert \sigma \rangle  \nonumber \\
&=&\sum_{\sigma \sigma '} \sum_{ij}   \varphi_{i}^{\ast} (\vec{r} ' \sigma ' q) \, \langle \sigma' \lvert \vec{\sigma} \rvert \sigma \rangle \,  \varphi_{j} (\vec{r} \sigma q) \,\rho^{q}_{ji} \,\,\, .
\label{vectorisoscalarden}
\end{eqnarray}
In the approximation that the single-particle wave-functions of spin-orbit partners are identical, it can be shown that the vector part of the density matrix $\vec{s}_{q} (\vec{r},\vec{r} ')$ is zero in spin-saturated nuclei.

\subsection{Long-term strategy}
\label{section:strategy}

Our long-term objective is to build so-called {\it non-empirical} nuclear energy functionals ${\cal E} [\rho]$ through the application of many-body perturbation theory implemented in terms of low-momentum interactions~\cite{Bogner:2003wn}
\begin{equation}
{\cal E} [\rho]= E^{HF} + \Delta E^{HF} \,\,\, ,
\end{equation}
where $E^{HF}$ denotes the (symmetry-unrestricted) Hartree-Fock (HF) contribution from two-, three-\ldots nucleon forces whereas $\Delta E^{HF}$ encompasses the corresponding correlation energy to all orders in perturbation theory\footnote{In applications to nuclei, except for doubly-magic ones, the ground-state energy will in fact be expanded around a quasi-particle vacuum of the Bogoliubov type rather than around a Slater determinant. This is necessary to take care of the Cooper pair instability that arises in the $^1S_0$ channel of the in-medium $NN$ amplitude.}. As opposed to the wisdom based on the use of conventional nuclear potentials, it has been shown recently that so-called low-momentum two- and three-nucleon interactions make the nuclear many-body problem more perturbative, with Hartree-Fock serving as a reasonable zeroth-order approximation~\cite{Bogner:2009un}. Still, calculations of the infinite nuclear matter (INM) equation of state~\cite{Bogner:2009un}, as well as binding energies and charge radii of doubly-magic nuclei~\cite{Roth:2005ah}, demonstrate that it is necessary to go at least to second-order in perturbation theory to resum enough bulk correlations into the EDF to get realistic binding. In the present paper though, we focus on the lowest-order contribution to the energy that is bilinear in the OBDM, i.e. the Hartree and Fock diagrams. While treating the direct (Hartree) term exactly, the objective of the density matrix expansion is to simplify the non-local character of the exchange (Fock) contribution to the energy by mapping it into a generalized Skyrme functional with density-dependent couplings. Therefore, the DME can be viewed as a constructive approach to encode finite-range physics into density-dependent couplings of a Skyrme-like functional.

The reasons for restricting our attention to the Hartree-Fock contributions in this initial study are two-fold. First, a non-trivial extension of the DME is needed to treat non-localities in both space and time that arise in higher orders of perturbation theory. I.e., one must properly account for the presence of energy denominators when designing a DME for 2nd-order MBPT and beyond. To date, a satisfactory generalization of the DME has not yet been formulated. Second, even if we follow the ad-hoc prescription of neglecting the non-locality in time by using averaged energy denominators, it is well established that the dominant contributions to bulk nuclear properties are of the Brueckner-Hartree-Fock (BHF) type. Operationally, this amounts to replacing the vacuum NN interaction in the Hartree-Fock expression by a Brueckner $G$-matrix (or a perturbative approximation in the case of low-momentum interactions) evaluated at some average energy. Since the $G$-matrix ``heals'' to the NN potential at long distances, applying the DME to the long-range part of the NN interaction at the Hartree-Fock level will in any event capture the same contributions to the density-dependent couplings as given by the long-range part of the $G$-matrix in a more sophisticated BHF calculation. In this way, the dominant density-dependence that arises from the finite-range of the inter-nucleon interactions is accounted for. Once a satisfactory generalization of the DME is developed to handle spatial and temporal non-locality on the same footing, non-localities arising from in-medium propagation can be mapped into the density-dependent Skyrme couplings as well.

\subsection{Two-nucleon interaction}
\label{section:interaction}

For simplicity, and because the main point of the present paper does not depend on it, we restrict our study to two-nucleon interactions only. Note however that a forthcoming publication is dedicated to the application of the presently developed DME to the HF energy derived from a chiral-EFT three-nucleon potential at N$^2$LO~\cite{gebremar09a}. In the present paper, we consider a generic local two-body interaction that
includes central, tensor and spin-orbit parts. Defining $x_i \equiv (\vec{r}_i \sigma_i q_i)$, one can write in the position $\otimes$ spin $\otimes$ isospin basis
\begin{eqnarray}
 \, \langle x_1  x_2 \lvert \,  {V}^{ST}_{I} \, \rvert x_3 x_4 \rangle &\equiv& V^{ST}_{I}
\, \delta (\vec{r}_1-\vec{r}_3) \, \delta (\vec{r}_2-\vec{r}_4) \,\,\, ,
\end{eqnarray}
where $I$ can be $C-$central,
$LS-$spin-orbit or $T-$tensor whereas $(S,T)$ takes values $(1,0), (0,1), (1,1)$ or $(0,0)$, where the first number 1/0 refers to two-body spin-triplet/singlet channels whereas the second number 1/0 refers to two-body isospin-triplet/singlet channels. More explicitly, the
central part of the interaction reads
\begin{eqnarray}\nonumber
V^{ST}_C &\equiv& v^{ST}_C(r)\,\,\Pi^{\sigma}_{s/t}\,\Pi^{\tau}_{s/t} \,\,\, ,
\end{eqnarray}
where the relative and center of mass coordinates are
defined as
\begin{eqnarray}
\vec{r}\equiv\vec{r}_1-\vec{r}_2 \, \,\,\text{and}\,\,\,  \vec{R}\equiv \frac{1}{2}(\vec{r}_1 + \vec{r}_2).
\end{eqnarray}
while spin/isospin
singlet/triplet projectors
\begin{eqnarray}
\Pi^{\sigma}_{s/t} \equiv \frac{1}{2} (1 -\!\!/\!\!+ P^{\sigma}_{12}) \, \,\,\text{and}\,\,\,  \Pi^{\tau}_{s/t} \equiv \frac{1}{2} (1 -\!\!/\!\!+ P^{\tau}_{12})\,,
\end{eqnarray}
are expressed in terms of spin/isospin exchange operators
\begin{eqnarray}
P^{\sigma}_{12} \equiv \frac{1}{2} (\sigma_1.\sigma_2 + 1) \, \,\,\text{and}\,\,\,  P^{\tau}_{12} \equiv  \frac{1}{2} (\tau_1.\tau_2 + 1)
\,.
\end{eqnarray}
The spin-orbit and tensor parts of the two-nucleon interaction take the form
\begin{eqnarray}
V^{ST}_{LS} &\equiv& -\frac{i}{2} \, v^{ST}_{LS}(r)\,\vec{r} \times
\vec{\nabla} \cdot (\vec{\sigma}_1 + \vec{\sigma}_2)
\,\,\Pi^{\sigma}_{s/t}\,\Pi^{\tau}_{s/t} \nonumber\,, \\
V^{ST}_{T} &\equiv& v^{ST}_{T}(r)\biggl[ 3 \bigl(\vec{\sigma}_1 \cdot
\vec{e}_r \bigr)\bigl(\vec{\sigma}_2 \cdot \vec{e}_r \bigr)
-\vec{\sigma}_1 \cdot \vec{\sigma}_2
\biggr]\,\Pi^{\sigma}_{s/t}\,\Pi^{\tau}_{s/t} \, , \nonumber
\end{eqnarray}
with
$\vec{e}_r \equiv \vec{r}/r$. It should be noted that the spin-orbit and
tensor parts of the interaction only act in the spin-triplet
channel.

\subsection{Fock contribution to the energy}

As mentioned earlier, the strategy consists of applying the DME to the exchange part of the HF energy while treating the Hartree term exactly. Indeed, it was realized long ago, starting with the early works on the DME by Negele and Vautherin~\cite{negele72,negele75}, that treating the direct part exactly has the following advantages:
\begin{itemize}
\item[(i)] It provides a better reproduction of the density fluctuations and the energy produced from an exact HF calculation~\cite{negele75}.
\item[(ii)] It significantly reduces the self-consistent propagation of errors if one restricts the DME to the exchange contribution~\cite{negele75,sprung75}.
\item[(iii)] There is no additional complexity in the numerical solutions of the resulting self-consistent HF equations~\cite{negele75} compared
to applying the DME to both Hartree and Fock terms.
\end{itemize}

The Fock contributions from central, spin-orbit and tensor
parts of the two-body interaction take the form
\begin{eqnarray}
E^{F}_C[ST] &\sim&\int d\vec{r}_1 d\vec{r}_2 \,
 \Big[ \rho_q (\vec{r}_1,\vec{r}_2) \, \rho_{q'}
(\vec{r}_2,\vec{r}_1) \nonumber \\
&&\hspace{-.0cm} +\,\,\,\vec{s}_q (\vec{r}_1,\vec{r}_2) \cdot \vec{s}_{q'}
(\vec{r}_2,\vec{r}_1) \Big] \,v^{ST}_C(r) \,\,, \label{central-contribution} \\[.25cm]
E^{F}_{LS}[ST] &\sim&  \int d\vec{r}_1 d\vec{r}_2 \,
 \Big [\rho_q (\vec{r}_1,\vec{r}_2) \, \vec{r} \times
\vec{\nabla}_2 \cdot \vec{s}_{q'}
(\vec{r}_2,\vec{r}_1) \nonumber \\
 &&\hspace{-.5cm} +\,\,\,\vec{s}_q (\vec{r}_1,\vec{r}_2)\cdot \vec{r}\times
\vec{\nabla}_2 \rho_{q'}
(\vec{r}_2,\vec{r}_1)\Big]\,v^{ST}_{LS}(r) , \label{spin-orbit-contribution}\\[.25cm]
E^{F}_T[ST] &\sim& \int d\vec{r}_1 d\vec{r}_2 \,
 \Big[\vec{s}_q (\vec{r}_1,\vec{r}_2)\cdot \vec{s}_{q'}
(\vec{r}_2,\vec{r}_1) \nonumber \\
&&\hspace{-0.85cm} +\, \sum_{\mu \nu} \frac{r_\mu r_\nu}{r^2} s_{q,\mu}
(\vec{r}_1,\vec{r}_2) s_{q',\nu} (\vec{r}_2,\vec{r}_1)\Big]\,v^{ST}_T(r) , \label{tensor-contribution}
\end{eqnarray}
where numerical coefficients and
overall signs, as well as sums and/or selection rules over isospin projections have been omitted. Indeed, only the {\it structure} of the terms at play is of importance for the present paper. For time-reversal invariant systems,
the scalar and vector parts of the OBDM satisfy the relations~\cite{engel75}
\begin{eqnarray}
\rho_q (\vec{r}_1 , \vec{r}_2 )& =&\rho_q( \vec{r}_2, \vec{r}_1) \label{eqn:scalarTR}\,\,\,,\\
\vec{s}_q (\vec{r}_1 , \vec{r}_2)& =& - \vec{s}_q (\vec{r}_2 , \vec{r}_1)\label{eqn:vectorTR}\,\,\, ,
\end{eqnarray}
such that the exchange contribution from the spin-orbit
interaction reduces to
\begin{eqnarray}\nonumber
E^{F}_{LS}[ST] &\sim&  \int d\vec{r}_1 d\vec{r}_2 \,
v^{ST}_{LS}(r) \, \vec{s}_q (\vec{r}_1,\vec{r}_2) \cdot \vec{r} \times
\vec{\nabla}_2 \rho_{q'}
(\vec{r}_2,\vec{r}_1) \,\,.
\end{eqnarray}

\section{Revisiting the DME}
\label{PSA-refomulation}

\subsection{Basics of the DME}
\label{subsection:basicsofdme}

The DME was originally proposed by Negele and Vautherin to establish a theoretical connection between the empirical zero-range Skyrme force and Hartree-Fock calculations with realistic NN interactions~\cite{negele72}. The central idea is to factorize the non-locality of the OBDM by expanding it into a finite sum of terms that are separable in relative and center of mass coordinates. Adopting notations similar to those introduced in
Refs.~\cite{doba03b} and ~\cite{doba05book}, one writes
\begin{eqnarray}
\rho_{q} (\vec{r}_1, \vec{r}_2)  &\approx&  \sum^{n_{\text{max}}}_{n=0} \Pi^{\rho}_n (k \, r) \,\, {\cal P}_n (\vec{R}) \,\,\, , \label{approxscalar}\\
\vec{s}_q (\vec{r}_1, \vec{r}_2) &\approx& \sum^{m_{\text{max}}}_{m=0} \Pi^{\vec{s}}_m (k\, r) \,\, {\cal Q}_m (\vec{R}) \,\,\, ,\label{approxvector}
\end{eqnarray}
where $k$ is a momentum scale to be determined that sets the scale for the decay in the off-diagonal direction, $\Pi^f_n (k \, r)$ are the so-called
$\Pi-$functions that remain to be specified, and $\{{\cal P}_n
(\vec{R}), {\cal Q}_m (\vec{R})\}$ denote various bilinear products of local densities and their gradients $\{\rho_q(\vec{R}), \tau_{q}(\vec{R}), J_{q,\mu\nu}(\vec{R}), \vec{\nabla} \rho_q (\vec{R}),
\Delta \rho_q(\vec{R})\}$ obtained from the OBDM through
\begin{eqnarray}
\rho_{q}(\vec{R})  &\equiv& \rho_q (\vec{r}_1, \vec{r}_2) |_{\vec{r}_1=\vec{r}_2=\vec{R}} \,\,\, , \\
\tau_{q}(\vec{R}) &\equiv&  \nabla_1 \cdot \nabla_2 \, \rho_q (\vec{r}_1, \vec{r}_2) |_{\vec{r}_1=\vec{r}_2=\vec{R}} \,\,\,,\\
J_{q,\mu \nu}(\vec{r}) &\equiv& - \frac{i}{2} (\nabla_1 - \nabla_2)_\mu \; s_{q, \nu} (\vec{r}_1,\vec{r}_2) |_{\vec{r}_1=\vec{r}_2=\vec{R}} \,\,\, .
\end{eqnarray}
The above local densities relate to the matter density, the kinetic density and the cartesian spin-current pseudotensor density, respectively. See Appendix~\ref{appendix:local-densities} for more details. Provided that large enough $n_{\text{max}}$ and $m_{\text{max}}$ give an accurate reproduction of the Fock contributions to the energy (Eqs.~\ref{central-contribution}, \ref{spin-orbit-contribution} and~\ref{tensor-contribution}), the benefit of expansion~\ref{approxscalar}-\ref{approxvector} is to provide a local approximation of the form (for time-reversal invariant systems)
\begin{widetext}
\begin{alignat}{1}\label{eqn:skryme-like-EDF}
E^{F} \approx &\sum_{q}\, \int  d \vec{R}\,\biggl\{
{A}^{\rho\rho} \,\rho_{q}(\vec{R}) \, \rho_{q}(\vec{R}) \, + \, {A}^{\rho
\tau} \, \rho_q(\vec{R}) \, \tau_{q}(\vec{R}) \, + \, {A}^{\rho \Delta \rho} \, \rho_q(\vec{R})
\,\Delta\, \rho_{q}(\vec{R}) \,
+ \, {A}^{\rho \nabla J} \, \rho_q(\vec{R}) \, \vec{\nabla}\,
\cdot \vec{J}_{q}(\vec{R}) \, \nonumber\\
& \hspace{2cm} +  \,{A}^{\nabla \rho
 J} \, \vec{\nabla}\rho_q(\vec{R}) \,  \cdot \vec{J}_{q}(\vec{R}) \,
 + \, {A}^{JJ} \,\sum_{\mu \nu} J_{q, \mu \nu} (\vec{R})\,J_{q, \mu \nu} (\vec{R})\nonumber\\
& \hspace{2cm}
+ \, {A}^{J\bar{J}} \,\biggl[ \biggl( \sum_{\mu } J_{q, \mu \mu} (\vec{R})\biggr) \biggl(\sum_{\mu } J_{q, \mu \mu} (\vec{R})\biggr)
\,+\,\sum_{\mu \nu}\, J_{q, \mu \nu} (\vec{R})\,J_{q, \nu \mu} (\vec{R})\biggr]\,
\biggr\} \nonumber \\
+ & \,\sum_{\bar{q}}\, \int  d \vec{R}\,\biggl\{ {B}^{\rho\rho}
\,\rho_{q}(\vec{R}) \, \rho_{\bar{q}}(\vec{R}) \, + \, {B}^{\rho \tau} \,
 \rho_q(\vec{R}) \, \tau_{\bar{q}}(\vec{R}) \, + \, {B}^{\rho \Delta \rho} \,
\rho_q(\vec{R}) \,\Delta\, \rho_{\bar{q}}(\vec{R}) \,+\, {B}^{\rho \nabla J} \, \rho_q(\vec{R}) \, \vec{\nabla}\,
\cdot \vec{J}_{\bar{q}}(\vec{R})\,\nonumber\\
& \hspace{2cm}
+ \,{B}^{\nabla \rho
 J} \, \vec{\nabla} \rho_q(\vec{R}) \,  \cdot \vec{J}_{\bar{q}}(\vec{R}) \,
+ \, {B}^{JJ} \,\sum_{\mu \nu} J_{q, \mu \nu} (\vec{R})\,J_{\bar{q}, \mu \nu} (\vec{R})
\nonumber\\
& \hspace{2cm} + \, {B}^{J\bar{J}} \,\biggl[ \biggl( \sum_{\mu } J_{q, \mu \mu} (\vec{R})\biggr) \biggl(\sum_{\mu } J_{\bar{q}, \mu \mu} (\vec{R})\biggr)
\,+\,\sum_{\mu \nu}\, J_{q, \mu \nu} (\vec{R})\,J_{\bar{q}, \nu \mu} (\vec{R})\biggr]\,
\biggr\}\,,
\end{alignat}
\end{widetext}
which is nothing but a local Skyrme-like EDF with couplings microscopically derived from the vacuum interaction. The couplings depend on the yet to-be-specified momentum scale $k$, and are given by
integrals of the finite-range NN interaction over various combinations of $\Pi$-functions, e.g.
\begin{equation}
\label{eq:schematiccoupling}
A^{\rho\rho}[k]  \sim 4\pi \int \! \! r^2 dr \, v^{ST}_C(r) \, \Big[\Pi^{\rho}_0(k\, r)\Big]^2  \,\,\, .
\end{equation}
Complete formulas for all the couplings appearing in Eq.~\ref{eqn:skryme-like-EDF} are provided in appendix~\ref{appendix:explicit-coupling-forms}. Before coming to the details of the DME method, a few remarks are in order:

\begin{itemize}
\item[(i)] Eventually, the momentum scale $k$ will be linked to the local Fermi momentum $k^q_F(\vec{R})$, or to a similar function, such that {\it all} couplings become density/position dependent. From Eq.~\ref{eq:schematiccoupling}, one sees that such density/position dependence is a direct consequence of the finite-range of the NN interaction. In this respect, the form given in Eq.~\ref{eqn:skryme-like-EDF} is more general than any existing empirical Skyrme EDF.
\item[(ii)] Due to such a density/position dependence of the couplings, terms that are usually connected through a partial integration, e.g. $\rho_q(\vec{R})
\,\Delta\, \rho_{q}(\vec{R})$ and $\vec{\nabla} \rho_q(\vec{R})  \cdot \vec{\nabla} \rho_{q}(\vec{R})$, can in general no longer be transformed into one another.
As a result, one keeps both types of terms explicitly in the resulting EDF.
\item[(iii)] Starting from a realistic vacuum Hamiltonian containing a three-nucleon force, one obtains a richer EDF including a wealth of trilinear terms~\cite{gebremar09a}. Including such terms will be eventually essential to any realistic application of the present work.
\item[(iv)] Eq.~\ref{eqn:skryme-like-EDF} is to be complemented with the Hartree contribution that can either be put under the form of a local EDF or treated exactly. Regardless, the EDF thus obtained only contains the physics of the HF approximation such that further correlations must be added in order to produce any reasonable description of nuclei. In the short term, such an addition can be done empirically by adding the above DME coupling functions to empirical Skyrme functionals and performing a refit of the Skyrme constants to data. This phenomenological procedure is motivated by the earlier observation that a Brueckner $G$-matrix differs from the vacuum NN interaction only at short distances.  Therefore, one can interpret the refit to data as approximating the short-distance part of the $G$-matrix with a zero-range expansion thru second order in gradients.  Eventually though, and as already stated, it is the goal of a future work to design a generalized DME that is suited to higher orders in perturbation theory.
\end{itemize}

\subsection{Existing variants of the DME}
\label{subsection:variantsofdme}

Several DME variants applicable to the HF energy have been developed in the past~\cite{negele72,campi77,meyer86,soubbotin99}. They mainly differ regarding (i) the choice made to fix the momentum scale $k$, (ii) the path followed to obtain actual expressions of
the $\Pi-$functions (see below) and (iii) the set of local densities that occur in the
expansion. For instance, the DME of Ref.~\cite{campi77} is a variant of the original one proposed by Negele and Vautherin (NV-DME)~\cite{negele72} that improves the accuracy of the expansion obtained at first order ($n_{\text{max}}=0$) by optimizing the momentum scale $k$.
The DME of Ref.~\cite{soubbotin99} is based
on a semi-classical extended Thomas-Fermi approximation, while the one proposed in
Ref.~\cite{meyer86} is a phenomenological method that introduces parameters to be optimized in order to obtain the correct local semiclassical
kinetic energy density and integrated projector identity of the OBDM (see Eq.~\ref{constraintN}).

\subsection{Motivation for a PSA reformulation of the DME}

The central part of the present work relates to a new and more general DME variant that is based on phase-space averaging (PSA) techniques. It will be denoted as PSA-DME throughout. The need for such a new formulation of the DME, in light
of the number of already available variants, relies on the following observations
\begin{itemize}
\item[(i)]
Existing DME formulations have focused mostly on the scalar part of the OBDM. For instance, Negele and Vautherin acknowledge in their seminal paper that they were not able to design an approximation of the vector part of the OBDM on the same level, and thus with the same accuracy, as the one they obtained for the scalar part. This is an essential problem in view of constraining non-empirically the nuclear EDF. Indeed, the vector part of the OBDM is non zero in spin-unsaturated nuclei, i.e. in almost all nuclei.
\item[(ii)] The PSA reformulation proposed below provides a consistent derivation of the
DME expansion of both the scalar and the vector pieces of the OBDM. In addition, it recovers the NV-DME as a particular case, such that one is offered the freedom to choose in a consistent fashion the variant that best optimizes the reproduction of the each of the three Fock contributions to the energy.
\item[(iii)]
In the PSA approach, one uses information from the local momentum phase space
distribution of the system of interest in order to optimize the
DME length-scale $k$ and to produce analytical expressions for the
$\Pi^f_n(k r)$ functions.
\item[(iv)]
Finally, it should be pointed out that all available DME
techniques hold only for time-reversal invariant systems. Hence, an
approach that can be extended to non time-reversal invariant systems
is important to constrain the nuclear EDF for non-time reversal invariant systems. In that respect,
the requirements of Galilean, alternatively gauge invariance, can be used to establish various relations between
the $\Pi-$functions multiplying certain time-even and time-odd densities~\cite{doba03b, gebremar08}.
\end{itemize}

Note that the PSA formulation of the DME is not completely new. Negele and Vautherin mentioned the possibility to use such an approach, having in mind to use the phase space of infinite nuclear matter, before reverting to a formal Bessel-function plane-wave expansion. From a formal point of view, the PSA approach developed below differs from that mentioned in Ref.~\cite{negele72} and is applied consistently to both the scalar and the vector parts of the OBDM. For
instance, in spite of the weak angular dependence of the scalar part
of the OBDM~\cite{martorell84}, the inconsistency in the order of application of
the angle-averaging and series expansion that exists in Ref.~\cite{negele72}
is not an issue in the present case.

\begin{figure}[hptb]
{\includegraphics[keepaspectratio,angle = 0,width=0.9\columnwidth]%
{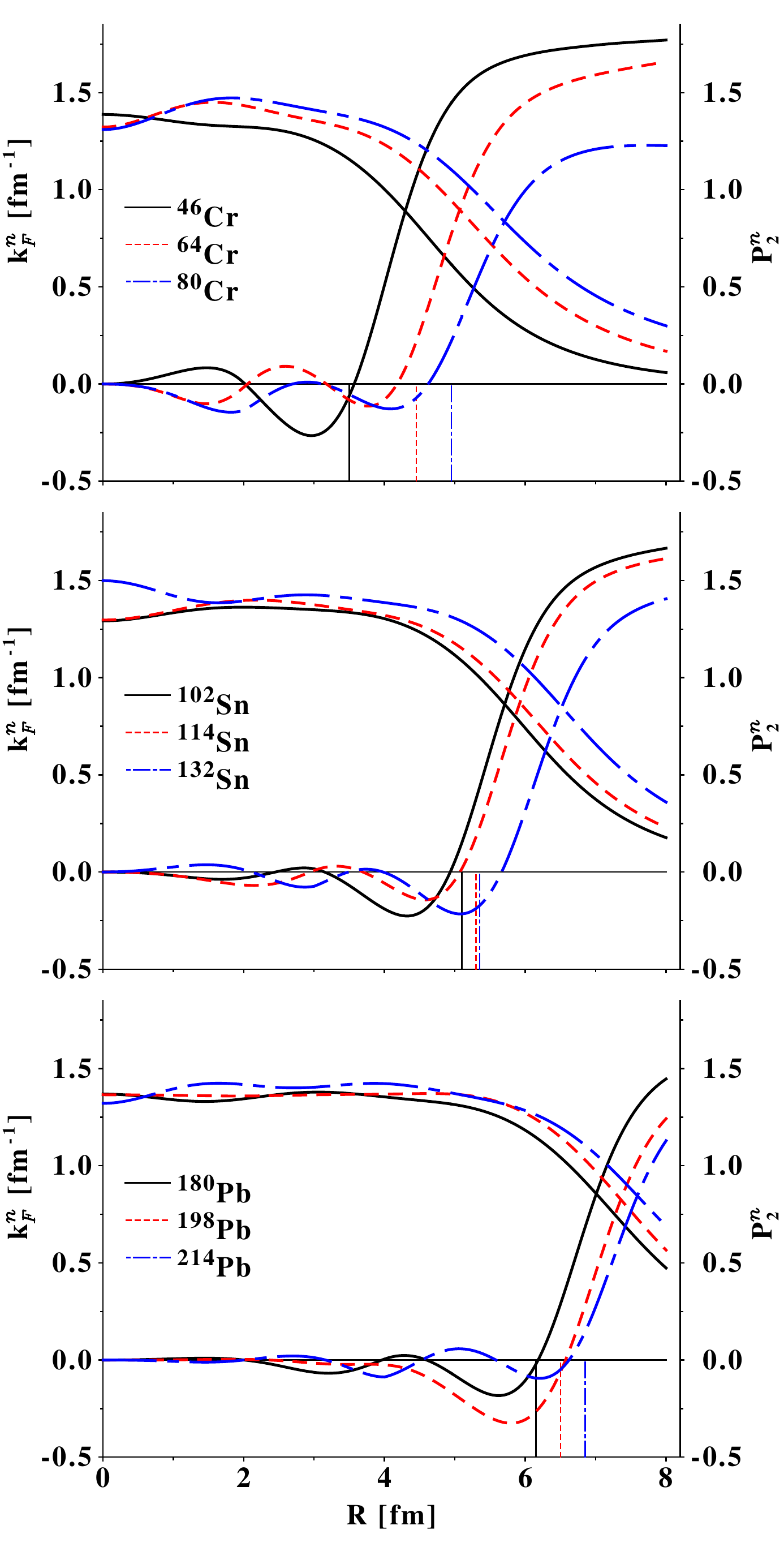}}\caption{ \label{fig:local_anisotropty} (Color online) The quadrupole anisotropy $P^n_2(\vec{R})$ of the local neutron momentum distribution in a
selected set of semi-magic nuclei. The black, red and blue vertical lines indicate
the approximate half-radii (where the density becomes half of the density at the origin).}
\end{figure}

\subsection{Momentum phase-space of finite Fermi systems}
\label{subsection:FFSphase-space}

A finite fermi system exhibits peculiar properties for the momentum phase-space distribution that are not present for homogeneous systems. The intent of this section is to mention those features that are relevant to the present work. The local momentum distribution of quantum systems can be studied via a multitude of quantum phase-space distribution functions~\cite{lee03}.
Using the Wigner distribution in Ref.~\cite{durand82} and the Husimi
distribution in Ref.~\cite{bulgac96}, the local single-particle momentum distribution is shown to display a diffuse and anisotropic Fermi surface when sitting at the (spatial) surface of the finite system. For
reasons discussed in Sec.~\ref{subsection:scalar-application}, the diffuseness is not as important as the anisotropy. Hence, we now describe a method that can be used to quantify of the
anisotropy of the local Fermi surface.

In Ref.~\cite{bulgac96}, the local quadrupolar deformation of the momentum Fermi surface (for a given isospin) is given by\footnote{As the anisotropy is usually not large, it is not necessary (at least in this work) to go to higher
multipoles to quantify the deformation.}
\begin{widetext}
\begin{eqnarray}
P^q_2(\vec{r})\,&\equiv&\,\frac{\int d\vec{p} \bigl[3(\vec{e}_r \cdot
\vec{p})^2 -\vec{p}^2 \bigr] H_q(\vec{r},\vec{p})}{\int d \vec{p} \,
\vec{p}^2 H_q(\vec{r},\vec{p})} = \biggl[\frac{3}{\tau_q(\vec{r})} \sum_{i} |(\vec{e}_r \cdot \vec{\nabla})
\varphi_i (\vec{r} q)|^2 \,  \rho^q_{ii} -1\biggr] + \mathcal{O}((k^q_F r_0)^2)\,, \label{eqn:P_2definition}
\end{eqnarray}
\end{widetext}
where $H_q(\vec{r},\vec{p})$ is the Husimi distribution,
$r_0$ is a length scale used in the Husimi
distribution and $k^q_F$ is a short-hand notation for the local Fermi momentum $k^q_F(\vec{R})$ defined in a local density approximation through
\begin{equation}\label{eqn:kFdef}
k^q_F \equiv \Bigl[3\, \pi^2 \,  \rho_q(\vec{R})\Bigr]^{1/3}\,\,\,.
\end{equation}
Equation~\ref{eqn:P_2definition} is computed in the basis $\varphi_i (\vec{r} q)$ that diagonalizes $\rho$, i.e. the basis from which the Slater determinant $| \Phi \rangle$ is built\footnote{When using a reference state of the Bogoliubov type, the corresponding basis of interest is the so-called canonical basis.}. A simplified expression of $P^q_2(\vec{r})$ in spherical symmetry suitable for semi-magic nuclei is provided in appendix~\ref{appendix:P_2derivation}.

Fig.~\ref{fig:local_anisotropty} shows the quadrupole anisotropy of the local neutron momentum distribution calculated for a selection of semi-magic nuclei. Single-particle wave-functions are obtained from a Skyrme-EDF calculation performed with the BSLHFB code~\cite{lesinski} using the
SLy4 parametrization of the Skyrme EDF with no pairing. Figure~\ref{fig:local_anisotropty} also displays the local neutron Fermi momentum (Eq.~\ref{eqn:kFdef}) in order to locate the position of the nuclear surface. In spite of pronounced shell fluctuations, the result corroborates the conclusions drawn in Ref.~\cite{bulgac96}; $P^n_2(\vec{R})$ becomes negative just inside the surface, denoting an oblate momentum Fermi surface while, outside this region, the local momentum Fermi surface becomes strongly prolate. In both cases, we have taken an axis normal to the nuclear surface as the reference axis. The next two sections show how we make use of these properties of the phase-space distribution of finite Fermi systems to design our PSA-DME.

\subsection{The scalar part of the OBDM}
\label{subsection:scalar-application}

In a nutshell, the PSA approach consists of three basic steps: (i) the isolation of the non-locality as an exponential derivative operator acting on the OBDM, (ii) the expansion of that operator around a momentum scale $\vec{k}$ and (iii) the averaging of that momentum scale over the local momentum distribution of the system of interest.

Applying the first two steps to the scalar part of the OBDM of a time-reversal invariant
system, one writes
\begin{widetext}
\begin{alignat}{3}
\rho_{q} \bigl(\vec{R} +
\frac{\vec{r}}{2},\vec{R}-\frac{\vec{r}}{2}
 \bigr)=&  \sum_{i \sigma} \varphi^\ast_{i } (\vec{r}_2  \sigma q) \, \varphi_{i }
(\vec{r}_1  \sigma q) \, \rho^q_{ii}
\nonumber\\
=& \, e^{i \vec{r} \cdot \vec{k}}\,e^{\vec{r} \cdot
\bigl(\frac{\vec{\nabla}_1 - \vec{\nabla}_2}{2} -i
\vec{k}\bigr)}  \, \sum_{i \sigma} \varphi^\ast_{i } (\vec{r}_2
\sigma q) \, \varphi_{i } (\vec{r}_1  \sigma q) \, \rho^q_{ii}
\bigg{|}_{\vec{r}_1=\vec{r}_2=\vec{R}} \, \nonumber\\
 \simeq  & \, e^{i \vec{r} \cdot \vec{k}}\,\biggl\{1 + \vec{r} \cdot
\biggl(\frac{\vec{\nabla}_1 - \vec{\nabla}_2}{2} -i \vec{k}\biggr) +
\frac{1}{2} \biggl[ \vec{r} \cdot \biggl(\frac{\vec{\nabla}_1 -
\vec{\nabla}_2}{2} - i \vec{k}\biggr)\biggr]^2\biggr\} \, \sum_{i \sigma} \varphi^\ast_{i } (\vec{r}_2 \sigma q) \, \varphi_{i }
(\vec{r}_1 \sigma q) \, \rho^q_{ii} \bigg{|}_{\vec{r}_1=\vec{r}_2=\vec{R}} \,\,\, .\label{eqn:second-order-expanded}
\end{alignat}
\end{widetext}
Before approximating the action of the non-locality operator, $e^{\vec{r} \cdot
(\vec{\nabla}_1 - \vec{\nabla}_2)/2}$, a
phase factor $e^{i \vec{r} \cdot \vec{k}}$ was extracted in order to perform a Taylor series expansion of the non-locality about the momentum scale $\vec{k}$. We presently truncate the expansion at second order although nothing prevents to study higher orders in principle. The next step consists in performing an angle averaging over the
orientation of $\vec{r}$, which is a reasonable step as the scalar
part of the OBDM has negligible dependence on the orientation of
$\vec{r}$~\cite{martorell84}. See appendix~\ref{appendix:scalar-derivation} for details.

The final step involves averaging the dependence on the momentum scale $\vec{k}$ over a model phase space that characterizes the system under study. Performing the PSA of a function $g(\vec{k})$ over the locally-equivalent pure isospin infinite matter phase-space, i.e. defining $G(\vec{k}^q_F)$ as
\begin{equation}\label{psa}
G(\vec{k}^q_F) \equiv \frac{3}{4\pi k^{q\, 3}_F} \int_{|\vec{k}| \leq k^q_F} \! d\vec{k} \, g(\vec{k})
\end{equation}
one obtains for time-reversal invariant systems
\begin{widetext}
\begin{eqnarray}
\rho_q (\vec{R} +  \frac{\vec{r}}{2},\vec{R} -
\frac{\vec{r}}{2} )& \simeq & \Pi^{\rho}_{0} (k^q_F r) \, \rho_q
(\vec{R})+\frac{r^2}{6}\Pi^{\rho}_{2} (k^q_F r)
\biggl[\frac{1}{4} \Delta \rho_q (\vec{R})- \tau_{q} (\vec{R}) +
\frac{3}{5} k^{q\,2}_F \rho_{q} (\vec{R})\biggr]\,, \label{DMEscalar}
\end{eqnarray}
\end{widetext}
with
\begin{eqnarray}
\Pi^{\rho}_{0} (k^q_F r)&\equiv& 3 \, \frac{j_1 (k^q_F(\vec{R}) r )}{k^q_F(\vec{R}) r} \label{pi-functions:scalar0}\,\,\,,\label{pirho0}\\
\Pi^{\rho}_{2} (k^q_F r)&\equiv& 3 \, \frac{j_1 (k^q_F(\vec{R}) r )}{k^q_F(\vec{R}) r} \label{pi-functions:scalar2}\,\,\, \label{pirho2}.
\end{eqnarray}
For details of the
derivation, refer to appendix~\ref{appendix:scalar-derivation}. Several comments are in order:
\begin{itemize}
\item[(i)] The phase space of finite nuclei has a marked difference from that of INM~\cite{durand82, bulgac96}. Still, using INM phase space suffices for the scalar part as will be apparent from the results
discussed in section~\ref{section:results-central}.
This is because, unlike the vector part of the OBDM discussed below, the scalar part is a bulk
quantity with most of its contribution coming from the interior of
the nucleus where, to a good approximation, the momentum distribution
resembles the one of INM~\cite{martorell84}.
\item[(ii)]  Dealing separately with the neutron or proton OBDM in a finite nucleus, it is natural to perform the corresponding PSA over the phase space of the locally-equivalent neutron or proton infinite matter. However, this provides $\Pi-$functions with an explicit isospin dependence that eventually breaks the explicit isospin invariance of the EDF (but not its isospin symmetry). Considering the small difference between $k^q_F$ and the total local momentum $k_F(\vec{R})$, defined in terms of the total density $\rho(\vec{R})\equiv\rho_n(\vec{R})+\rho_p(\vec{R})$ through
\begin{equation}\label{eqn:kFdeftotal}
k_F \equiv \Biggl[\frac{3\pi^2}{2} \,  \rho(\vec{R})\Biggr]^{1/3}\,\,\,,
\end{equation}
it might be preferred to perform the PSA over the phase space of symmetric nuclear matter, even in a neutron rich nucleus. In any case, all results presented below are obtained using $k^q_F$ but would not be significantly different if using $k_F$ instead.
\item[(iii)] The DME is not a naive Taylor expansion of the OBDM with respect to the non-locality $\vec{r}$. The $\Pi-$functions resum dependencies on $r$ to all orders such that the long distance limit behavior of the OBDM is reproduced (see below). However, as noted in Ref.~\cite{negele72}, the truncation of the expansion about $\vec{k}$ to second order leaves the specific value of the coefficients of terms beyond $k^q_F r$ undetermined (in the Taylor series expansion of $\Pi^{\rho}_{2} (k^q_F r)$). This indeterminateness gives one the freedom to optimize $\Pi^{\rho}_{2}$, which can be viewed as selecting a different rearrangement and truncation of the expansion~\cite{negele72}.
\item[(iv)] The zeroth-order $\Pi-$function $\Pi^{\rho}_{0} (k^q_F r)$ found above is exactly the one found in the original NV-DME of Ref.~\cite{negele72}. Just as in the NV-DME, the leading term of the PSA-DME reproduces the exact OBDM of infinite nuclear matter. The second order $\Pi-$function $\Pi^{\rho}_{2} (k^q_F r)$ is different\footnote{The Bessel expansion of Ref.~\cite{negele72} provides $\Pi^{\rho}_{2}= 105 \, j_3 (k^q_F r )/ (k^q_F r)^3$.} from the one found in Ref.~\cite{negele72}. However, this relates to the previous remark that emphasized the freedom in choosing the second-order $\Pi$-function. Moreover, we will find in Section~\ref{section:results} that these differences are rather small for contributions to the Fock energy. Therefore, our PSA-DME of the scalar part of the OBDM is essentially equivalent to the NV-DME of Ref.~\cite{negele72}.
\end{itemize}

The freedom mentioned above can be used to adjust
$\Pi^{\rho}_{2}$ to satisfy certain properties of
the exact OBDM, or simply to optimize the quality of the approximation through a comparison with realistic a OBDM. One example relates to the integrated idempotency of the OBDM, e.g. for neutrons
\begin{eqnarray}
N& =& \int d \vec{r} \,\rho_n (\vec{r}) = \int \int d\vec{r}_1 d \vec{r}_2 \,|\rho_n (\vec{r}_1,\vec{r}_2)|^2
\,\,\, . \label{constraintN}
\end{eqnarray}
As shown in Ref.~\cite{Koehl95},
there is a class of DME that satisfies this constraint. Unfortunately, the
$\Pi^{\rho}_{2}$ given in Eq. \eqref{pi-functions:scalar2} does not satisfy this
constraint. Even though the non-self consistent result given in~\ref{section:results-central} is satisfactory, this might not be the case in
a self-consistent test.

Other constraints on the $\Pi-$functions come from the expected limits for large and small values of
$r$. The $\Pi-$functions should go
to zero in the large $r$ limit, while for small $r$, the expansion must to reduce to a simple Taylor series. These requirements\footnote{The small $r$ constraints are obtained by setting
$r=0$ after the Taylor expansion is performed.} lead to~\cite{doba03b, gebremar08}
\begin{eqnarray}
    \Pi^{\rho}_0(0) &=& \Pi^{\rho}_2(0) = 1 \,\,\,, \\
    \Pi^{\rho \, \prime}_0(0) &=& \Pi^{\rho \,\prime \prime}_2 (0) \,\,\, ,  \\
    \lim_{\substack{r\rightarrow \infty}} \Pi^{\rho}_{0} &=& \lim_{\substack{r\rightarrow \infty}}
                        \Pi^{\rho}_{2} = 0 \,\,\, .
\end{eqnarray}
It can easily be shown that the above constraints are satisfied by the $\Pi-$functions
listed in Eqs.~\eqref{pi-functions:scalar0} and~\eqref{pi-functions:scalar2}.

\subsection{The vector part of the OBDM}
\label{subsection:vector-application}

Restricting again the discussion to time-reversal invariant systems and applying the same steps as for the scalar part of the OBDM, one obtains for its vector part
\begin{widetext}
\begin{alignat}{3}
\vec{s}_{q} \biggl(\vec{R}+ \frac{\vec{r}}{2} , \vec{R}-
\frac{\vec{r}}{2} \biggr)  =&  \sum_{i  \sigma_1 \sigma_2} \, \varphi^{\ast}_{i} (\vec{r}_2 \sigma_2 q) \,
\langle \sigma_2 \lvert \vec{\sigma} \rvert \sigma_1 \rangle \, \varphi_{i }
(\vec{r}_1 \sigma_1 q)  \, \rho^q_{ii}
\nonumber\\
=& \, e^{i \vec{r} \cdot \vec{k}}\,e^{\vec{r} \cdot
\bigl(\frac{\vec{\nabla}_1 - \vec{\nabla}_2}{2} -i
\vec{k}\bigr)}  \sum_{i  \sigma_1 \sigma_2} \, \varphi^{\ast}_{i} (\vec{r}_2 \sigma_2 q) \,
\langle \sigma_2 \lvert \vec{\sigma} \rvert \sigma_1 \rangle \, \varphi_{i }
(\vec{r}_1 \sigma_1 q)  \, \rho^q_{ii}
\bigg{|}_{\vec{r}_1=\vec{r}_2=\vec{R}} \, \nonumber\\
 \simeq  & \, e^{i \vec{r} \cdot \vec{k}}\,\biggl\{1 + \vec{r} \cdot
\biggl(\frac{\vec{\nabla}_1 - \vec{\nabla}_2}{2} -i \vec{k}\biggr) \biggr\}  \, \sum_{i  \sigma_1 \sigma_2} \, \varphi^{\ast}_{i} (\vec{r}_2 \sigma_2 q) \,
\langle \sigma_2 \lvert \vec{\sigma} \rvert \sigma_1 \rangle \, \varphi_{i }
(\vec{r}_1 \sigma_1 q)  \, \rho^q_{ii}  \bigg{|}_{\vec{r}_1=\vec{r}_2=\vec{R}} \,\,\, ,\label{eqn:first-order-expandedvector}
\end{alignat}
\end{widetext}
where only the first order term in the expansion of the non-locality operator was kept for reasons explained below. One also notes that the zero-order term provides the local spin density $\vec{s}_{q} (\vec{R})$ which is zero for the time-reversal invariant systems we are considering. In Ref.~\cite{negele72}, it was argued that averaging over the orientation of $\vec{k}$ and setting $k=k^q_F$ should be sufficient to provide a reasonable account of the vector part of the exact OBDM. This gives
\begin{equation}\label{basicvectorexpansion}
\vec{s}_{q,\nu} \biggl(\vec{R}+ \frac{\vec{r}}{2} , \vec{R}-
\frac{\vec{r}}{2}\biggr)  \simeq   i \,
\Pi^{\vec{s}}_1 (k^q_F r) \,\sum_{\mu} r_{\mu} J_{q, \mu \nu} (\vec{R}) \,,
\end{equation}
where
\begin{eqnarray}\label{pis1function_NV}
\Pi^{\vec{s}}_1 (k^q_F r) &=&j_0 (k^q_F(\vec{R}) r)\,\,\, .
\end{eqnarray}
If instead one applies the same procedure as for the scalar part of the OBDM and performs the PSA over the locally-equivalent pure-isospin infinite matter phase-space, one obtains\footnote{See appendix~\ref{appendix:vector-derivation} for details.}
\begin{eqnarray}\label{pis1function_NV}
\Pi^{\vec{s}}_1 (k^q_F r)\, &=& 3 \, \frac{j_1 (k^q_F(\vec{R}) r)}{k^q_F(\vec{R}) r }\,\,\, .
\end{eqnarray}

However, as mentioned in section~\ref{subsection:FFSphase-space}, the local momentum
distribution in the surface region of a finite nucleus has a markedly different behavior than the isotropic momentum distribution of infinite nuclear matter. Given that
the vector part of the density matrix peaks around the nuclear
surface, it seems more appropriate to perform the PSA over a deformed Fermi sea that incorporates the information contained in the function $P^q_2(\vec{R})$ discussed in section~\ref{subsection:FFSphase-space}. The details are given in appendix~\ref{appendix:vector-derivation}.
The final result differs from that in Ref.~\cite{negele72} only in the
analytical form of $\Pi^{\vec{s}}_1$. The result reads
\begin{eqnarray}\label{pis1function_PSA}
\Pi^{\vec{s}}_1 (\tilde{k}^q_F r) &=& 3 \, \frac{  j_1 (\tilde{k}^q_F(\vec{R}) r)}{\tilde{k}^q_F(\vec{R}) r}\,\,\,,
\end{eqnarray}
where
\begin{eqnarray}
\tilde{k}^q_F&\equiv&\biggl(\frac{2 + 2 \,P^q_2(\vec{R})}{2-P^q_2(\vec{R})}\biggr)^{1/3} k^q_F(\vec{R}) \,\,\,. \label{newscale}
\end{eqnarray}

The PSA over the locally-equivalent neutron or proton infinite matter modifies the analytical form of $\Pi^{\vec{s}}_1$ compared to NV-DME, i.e. compare Eqs.~\ref{pis1function_NV} and~\ref{pis1function_PSA}. In addition, and contrary to the scalar part of the OBDM for which it is unimportant, taking into account the deformation of the local momentum distribution of the finite system leads to a modification of the relevant momentum scale $\tilde{k}^q_F$. In view of isolating the significance of such an effect, while preserving the benefit of using PSA, one can set $P^q_2(\vec{R}) = 0$ in Eq.~\ref{newscale}. In Sec.~\ref{section:tensor-contribution}, we discuss and compare the accuracy obtained using all of the preceding variants.

Note that the expansion was limited to first order in Eq.~\ref{eqn:first-order-expandedvector}. The reason is that, for time-reversal invariant systems, the cartesian spin-current pseudotensor density $J_{q,\mu \nu} (\vec{R})$ and its gradients are the only standard local densities at hand to express the DME. Given that, we could not find any closed and parameter-free expression of higher-order contributions in terms of such local densities only. This points however to the possibility to study higher-order terms in the context of the generalized Skyrme EDF discussed in Ref.~\cite{carlsson09}.

Finally, one can easily verify that the large and small $r$ limits, viz,
\begin{equation}
\Pi^{\vec{s}}_1 (0) = 1\,\, , \,\,\Pi^{\vec{s}\, '}_1 (0)=0 \,\, \text{and} \,\,
\lim_{\substack{r\rightarrow \infty}} \Pi^{\vec{s}}_{0} =0 \,\,\,,
\end{equation}
mentioned at the end of section~\ref{subsection:scalar-application} are satisfied by the expressions of $\Pi^{\vec{s}}_1$ given by either Eq. \eqref{pis1function_NV} or Eq.\eqref{pis1function_PSA}.

\section{Comparing PSA- and NV-DME}
\label{section:results}

The accuracy of our newly developed PSA-DME needs to be tested against both non-self consistent and self-consistent HF calculations. A self-consistent test of the PSA-DME is the aim of a forthcoming publication. As explained below, we limit ourselves in the present paper to gauging the accuracy of the NV-DME and the PSA-DME against two non self-consistent measures.  Where relevant, we also set $P^q_2(\vec{r})=0$ in the PSA-DME of the vector part of the OBDM to isolate the significance of using a deformed local momentum Fermi surface. We denote that last variant as INM-DME.

\subsection{Inputs to non-self-consistent tests}
\label{section:schemforces}

The generic form of the central, spin-orbit and tensor interactions considered here have been given in Sec.~\ref{section:interaction}. The radial
form factors used in the present calculations for either of those interactions take the form (i) a gaussian or (ii) a renormalized Yukawa (according to Ref.~\cite{lepage97}). Specifically we use
\begin{equation}
v^{ST}_I(r) = \left\{
\begin{array}{l}
v_0 \,e^{-r^2/a^2}\,\,\,,  \\
\\
 \frac{v_0}{2 r} \bigl[e^{-m_\pi r}\text{erfc}\bigl(\frac{m_\pi}{\lambda} -r \lambda\bigr)-\bigl(r\rightarrow -r \bigr) \bigr]\,\,\,,
\end{array} \right.
\end{equation}
independently of the $(S,T)$ channel and with $v_0 = 50$ MeV, $a = 1.5$ fm, $m_\pi=0.7$ fm$^{-1}$. The momentum cut-off $\lambda$ is set equal to $2.1$ fm$^{-1}$ while
$\text{erfc}$ is the complementary error function. It must be stressed that none of these interactions are realistic two-nucleon interactions, but rather schematic representatives. The objective of the present study is to gauge the accuracy of the DME variants against a reasonable reference point that is not itself meant to provide useful or realistic results. The application of the present DME scheme to realistic chiral two- and three-nucleon interactions is the objective of a forthcoming publication~\cite{gebremar09a}. Finally, note that neutron density matrices and local densities used in the following sections have been obtained, for all semi-magic nuclei of interest, through spherical self-consistent EDF calculations employing the SLy4 EDF parameterizations with no pairing.

\subsection{Fock contribution from $V_C$}
\label{section:results-central}

The expression of the Fock contribution to the energy from the central part of the two-nucleon interaction is
given in Eq.\eqref{central-contribution}. It contains a bilinear product of
non-local matter densities as well as a bilinear product of non-local spin densities. Since the latter
also appears as part of the tensor contribution to the Fock energy (see Eq.\eqref{tensor-contribution}), we postpone the discussion
regarding the spin-density product to section \ref{section:tensor-contribution}.

Before comparing the Fock energy to its DME counterpart, we first conduct a more stringent test on the energy density in which the integration over the angle of $\vec{r}$ has already been performed, i.e. we compare the integrand
\begin{eqnarray}
C^F_{nn} (\vec{R},r)&\equiv& \frac{1}{4 \pi}\int \! d \vec{e}_r \, \rho_n(\vec{r}_1, \vec{r}_2) \,
\rho_n (\vec{r}_2, \vec{r}_1) \,\,\, , \label{eqn:CEprofile}
\end{eqnarray}
to its DME counterpart
\begin{widetext}
\begin{eqnarray}
C^{DME}_{nn}(\vec{R},r) &\equiv& \Bigl[\Pi^{\rho}_{0}(k^n_F r)\Bigr]^2 \, \rho_n
(\vec{R}) \, \rho_n (\vec{R}) + \frac{r^2}{3} \Pi^{\rho}_{0}(k^n_F r) \, \Pi^{\rho}_{2}(k^n_F r) \, \rho_{n} (\vec{R}) \biggl(\frac{1}{4} \Delta \rho_n (\vec{R})-\tau_{n} (\vec{R}) + \frac{3}{5} k^{n\, 2}_F \rho_{n} (\vec{R})\biggr)
 \, ,\label{eqn:CDprofile}
\end{eqnarray}
\end{widetext}
where the latter depends on which variant of the DME has been adopted\footnote{We denote such integrands as {\it energy densities} throughout the paper. Strictly speaking, it is necessary to multiply them by the interaction to obtain the dimension of an energy density. Still, we postpone the folding with the interaction to the second measure introduced below.}. Having in mind existing empirical Skyrme EDFs that contain only up to two spatial derivatives, terms containing fourth-order gradients have been truncated in $C^{DME}_{nn}(\vec{R},r)$. A consistent account of such fourth-order derivatives in the EDF would require to go also to fourth order in the DME itself, which is beyond the scope of the present study. This is an important point that underlines our philosophy that the primary purpose of the DME method is not to reproduce the fine details of the OBDM, but rather to reproduce as best as possible the energy density and the total energy at a given order in the expansion. The latter two are precisely what is gauged in this paper, whereas no tests dedicated to the reproduction of the OBDM by itself are performed.

\begin{figure}[hptb]
{\includegraphics[keepaspectratio,angle = 0,width=\columnwidth]%
{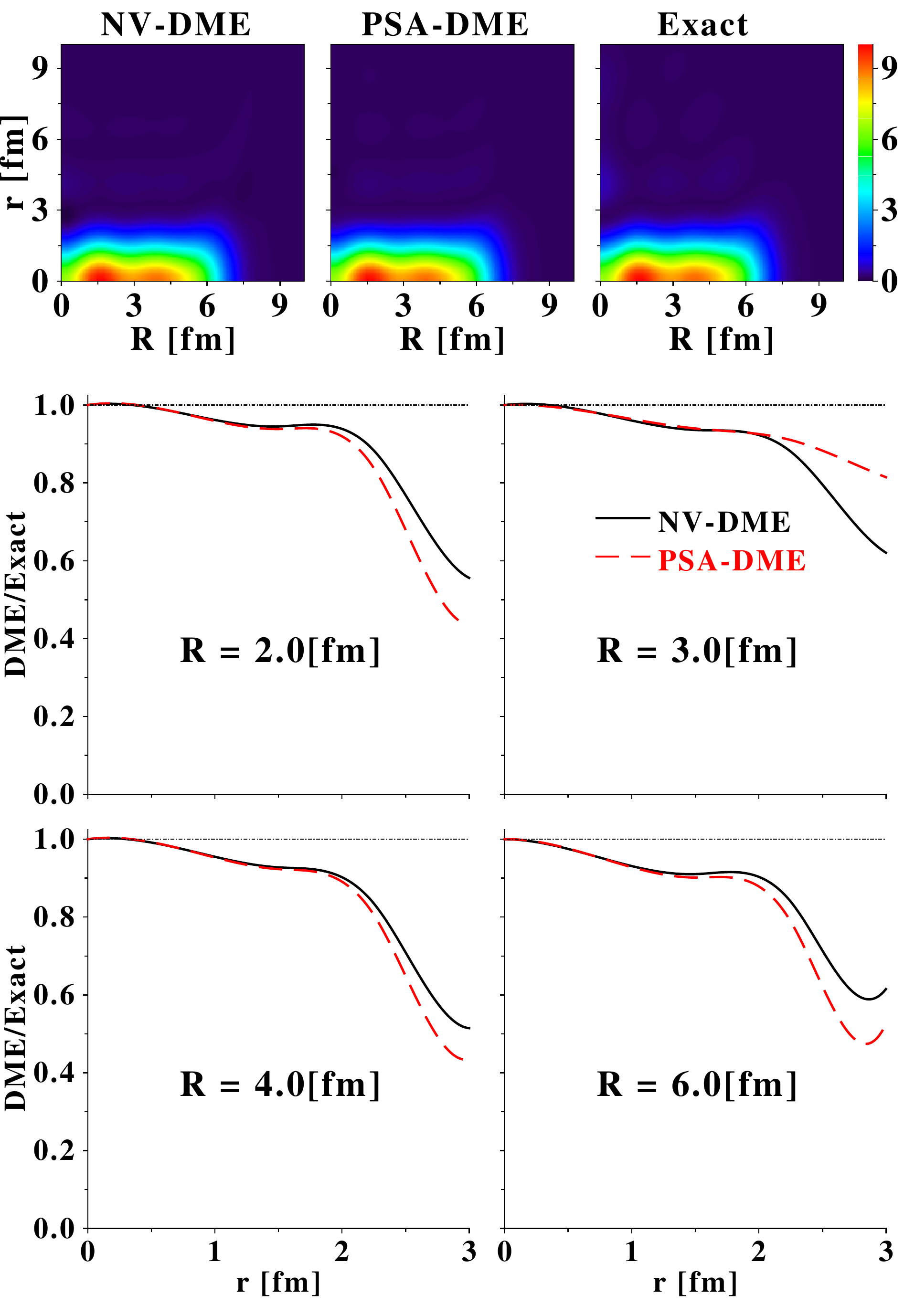}}\caption{ \label{fig:central-profile}(Color online) Comparison of $C^F_{nn}(\vec{R},r)$
and $C^{DME}_{nn}(\vec{R},r)$ where the latter is either computed from NV-DME or PSA-DME $\Pi-$functions. Upper panels: two-dimensional integrands. Lower panels:  ratios of $C^{DME}_{nn}(\vec{R},r)$ over $C^{F}_{nn}(\vec{R},r)$ for fixed values of $R$. Densities are obtained from a self-consistent EDF calculation of
$^{208}$Pb with the SLy4 Skyrme EDF in the particle-hole part and no pairing.}
\end{figure}

Figure~\ref{fig:central-profile} shows\footnote{Note that for semi-magic spherical nuclei used in the present paper, the energy densities $C^{F}_{nn}(\vec{R},r)$ and $C^{DME}_{nn}(\vec{R},r)$ only depend on the magnitude of $\vec{R}$.} that both NV-DME and PSA-DME provide comparably good profile-reproduction of the integrand $C^F(\vec{R},r)$ within the typical range of nuclear interactions
($r\!\sim\!2$ fm). Beyond such a non locality, the quality of the reproduction deteriorates significantly, with that of PSA-DME deteriorating slightly faster. In addition, one sees from the lower panels of Fig.~\ref{fig:central-profile} that the quality of the reproduction decreases as one goes to the nuclear surface, i.e. for $R\!\gtrsim\!4$ fm. This could be slightly improved by taking into account the deformation of the local momentum distribution when designing the PSA-DME for the scalar part of the OBDM,  which we do not do here. Note also that, although the plots are provided for two sample nuclei, more systematic tests have been performed over several semi-magic isotonic and isotopic chains that support such conclusions.

Coming to the energy itself, i.e. to the integrated product of the interaction $v_C(r)$ with the central energy density, we compare\footnote{We do not analyze individual couplings of the Skyrme-like EDF produced through the DME (Eq.~\ref{eqn:skryme-like-EDF}) in the present paper, but rather test the complete Fock energy provided by each of the terms (i.e. central, tensor, spin-orbit) of the two-nucleon interaction. We postpone to a forthcoming publication~\cite{gebremar09a} the analysis of the EDF couplings computed from realistic two- and three-nucleon chiral interactions using appendix~\ref{appendix:explicit-coupling-forms}.}
\begin{eqnarray}
E^F_C [nn]&=& 4 \pi \!\int\! d \vec{R} \,d r\, r^2 \, v_C(r)\, C^F_{nn}(\vec{R},r)\, ,
\label{eqn:VCE} \\
E^{DME}_C [nn] &=& 4 \pi \!\int\! d \vec{R} \,d r\, r^2 \, v_C(r)\, C^{DME}_{nn}(\vec{R},r).
\label{eqn:VCD}
\end{eqnarray}

\begin{figure}[hptb]
{\includegraphics[keepaspectratio,angle = 0,width=0.9\columnwidth]%
{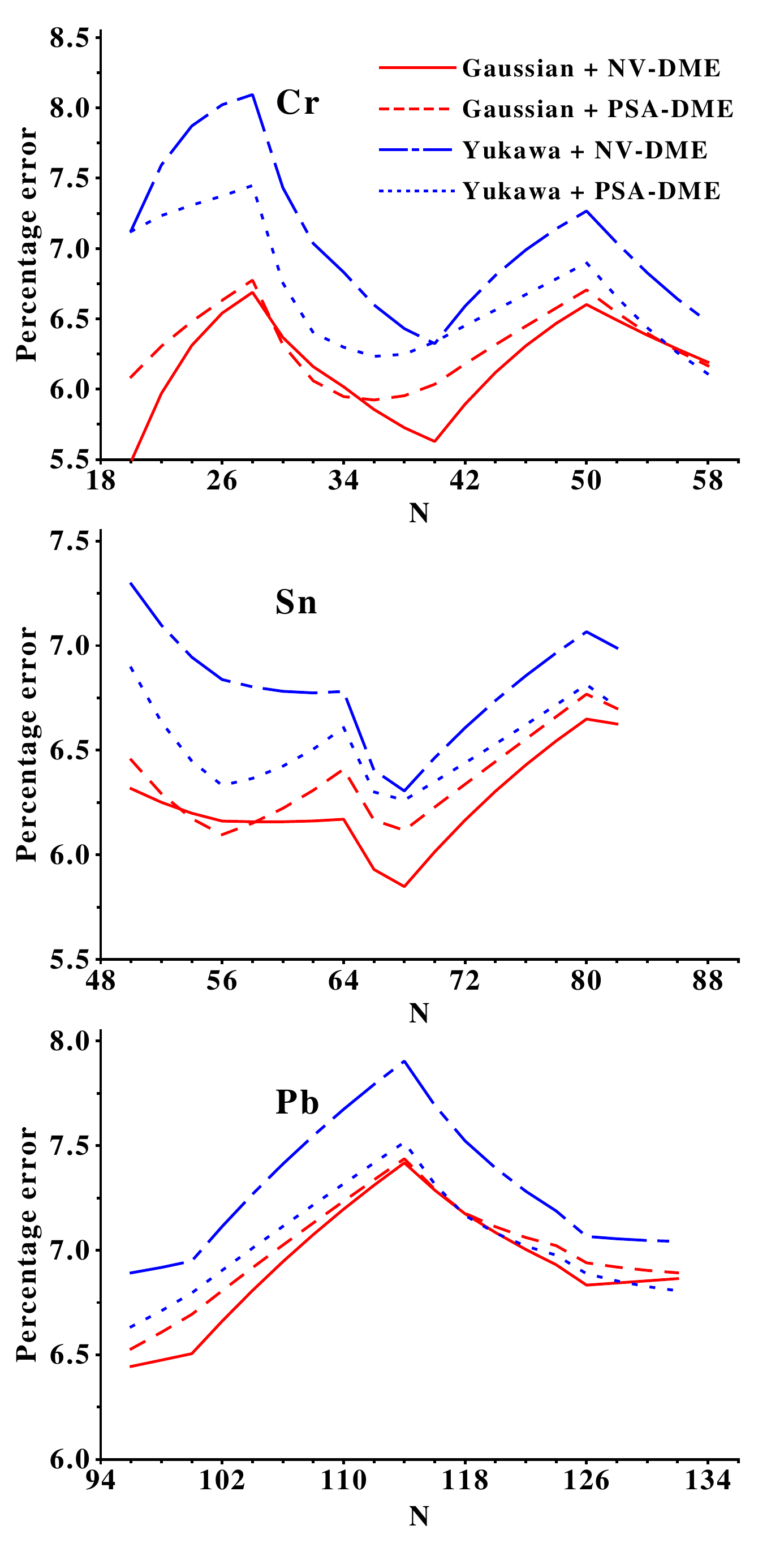}}\caption{ \label{fig:central-integrated}(Color online) Percentage error of $E^{DME}_C [nn]$ compared to $E^F_C [nn]$, where the former is either computed from NV-DME or PSA-DME $\Pi-$functions. Densities are obtained from self-consistent EDF calculations using the SLy4 Skyrme EDF in the particle-hole channel and no pairing.}
\end{figure}

\begin{figure}[hptb]
{\includegraphics[keepaspectratio,angle = 0,width=0.9\columnwidth]%
{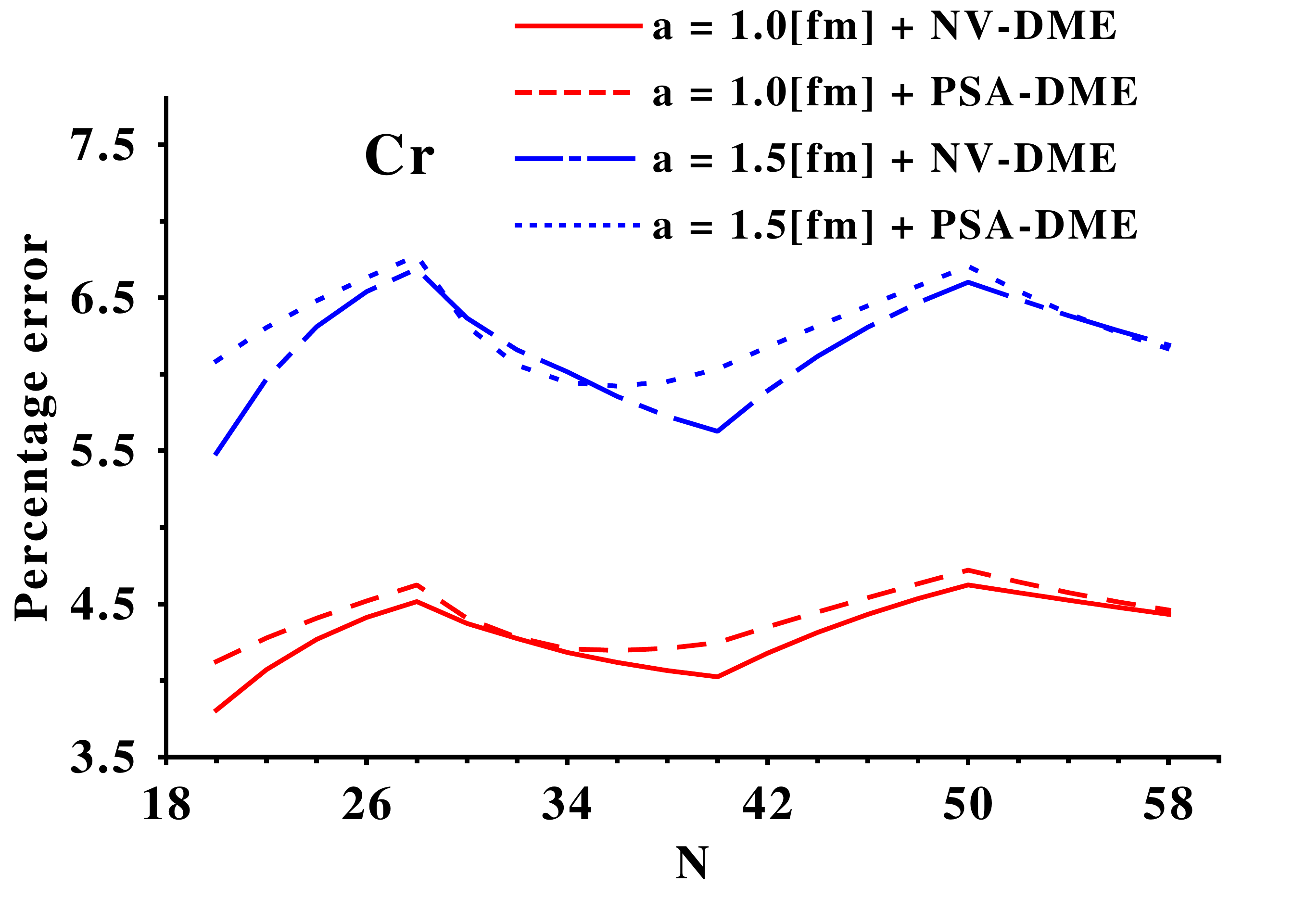}}\caption{ \label{fig:DME-RANGE}(Color online) The same as Figure \ref{fig:central-integrated} but for two different values of the range of the Gaussian interaction.}
\end{figure}
Figure~\ref{fig:central-integrated} shows the relative error obtained from the two DME variants compared to the exact Fock contribution for both the Gaussian and the renormalized-Yukawa radial form factors and for three semi-magic isotopic chains.

Let us start with Fig.~\ref{fig:DME-RANGE} that shows that the dependence of the accuracy on the
range of the (Gaussian) interaction used is significant, i.e. about a factor of two between $a=1.0$ fm and $a=1.5$ fm. As can be expected from the two-dimensional density profiles in Fig.~\ref{fig:central-profile}, the accuracy decreases as the range of interaction increases, which holds for all available DME techniques ~\cite{negele72,campi77,meyer86,soubbotin99}. This stresses that the local quasi-separability of the OBDM with respect to $\vec{r}$ and $\vec{R}$ underlining the DME, which is exact in INM, deteriorates with increasing non-locality $r$ in finite nuclei. As long as the hypothesis of quasi-separability is well realized within the range of the interaction, the DME can be quantitatively successful.

On average, the error obtained with PSA-DME and NV-DME are similar as can be seen in Fig.~\ref{fig:central-integrated}, i.e. about $6\!-\!8\%$ for the three isotopic chains and for both for the Gaussian and the renormalized-Yukawa interactions. In a future publication, we demonstrate that one can obtain a better accuracy (1-2\% error) by using a parameterized and empirically optimized phase-space distribution that takes the
diffuseness of the Fermi surface into consideration. A similar improvement over that of Ref.~\cite{negele72} is reported in Refs.~\cite{campi77,meyer86}.

\subsection{Fock contribution from $V_T$}
\label{section:tensor-contribution}

We now turn to the Fock contribution coming from the tensor part of the two-nucleon interaction. As shown by Eq.~\eqref{tensor-contribution}, such a contribution involves bilinear products of non-local spin densities. As a matter of fact, two terms with different analytical structures emerge such that the exchange tensor energy-density reads\footnote{We recall that the weights of the two terms have been omitted in agreement with Eq.~\ref{tensor-contribution}.}
\begin{eqnarray}
T^{F}_{nn}(\vec{R},r) &\equiv& T^{F}_{nn, 1}(\vec{R},r) \, + \, T^{F}_{nn, 2}(\vec{R},r) \,\,\,,\\
T^{F}_{nn, 1}(\vec{R},r) &\equiv& \frac{1}{4 \pi} \int \! d \vec{e}_r \, \vec{s}_n(\vec{r}_1, \vec{r}_2)
\cdot \vec{s}_n (\vec{r}_2, \vec{r}_1) \,\,\,,\\
T^{F}_{nn,2 }(\vec{R},r) &\equiv& \frac{1}{4 \pi} \int \! d \vec{e}_r  \,\sum_{\mu \nu}\,\frac{r_\mu r_\nu}{r^2}\,
s_{n,\mu}(\vec{r}_1, \vec{r}_2) \,\nonumber\\
&& \quad \quad \quad \quad \times \, s_{n,\nu}(\vec{r}_2, \vec{r}_1) \,\,\, ,
\end{eqnarray}
where $T^{F}_{nn, 1}(\vec{R},r)$ also appear in the central contribution to the Fock energy. The two DME counterparts, which eventually depend on which variants of the DME is being adopted, read
\begin{eqnarray}
T^{DME}_{nn,1}(\vec{R},r) &\equiv& - \frac{r^2}{3} \,\bigl[\Pi^{\vec{s}}_{1} (\tilde{k}^n_F r) \bigr]^2
\sum^{z}_{\mu,\nu = x} J_{n, \mu \nu}(\vec{R}) \, J_{n, \mu \nu}(\vec{R}) \, , \nonumber \label{eqn:TDprofile}\\
T^{DME}_{nn,2}(\vec{R},r) &\equiv& - \frac{r^2}{15} \,\bigl[\Pi^{\vec{s}}_{1} (\tilde{k}^n_F r) \bigr]^2
\sum^{z}_{\mu,\nu = x} \,\biggl(J_{n, \mu \nu}(\vec{R}) \, J_{n, \mu \nu}(\vec{R})\nonumber\\
&& +  J_{n, \mu \mu}(\vec{R})  J_{n, \nu \nu}(\vec{R}) + J_{n, \mu \nu}(\vec{R}) J_{n, \nu \mu}(\vec{R}) \biggr) ,\nonumber \label{eqn:TDprofile2}
\end{eqnarray}
and reduce for spherical systems to
\begin{eqnarray}
T^{DME}_{nn,1}(\vec{R},r) &\equiv& - \frac{r^2}{6} \,\bigl[\Pi^{\vec{s}}_{1} (\tilde{k}^n_F r) \bigr]^2
 \vec{J}_n(\vec{R}) \cdot \vec{J}_n(\vec{R}) \, ,\\
T^{DME}_{nn,2}(\vec{R},r) &\equiv& 0 \, .
\end{eqnarray}
One recovers a pattern which is seen when deriving the empirical Skyrme EDF from an auxiliary Skyrme effective interaction. That is, the central part of the interaction only produces the so-called {\it symmetric} bilinear tensor terms proportional to $J_{n, \mu \nu}(\vec{R}) \, J_{n, \mu \nu}(\vec{R})$ while $T^{DME}_{nn,2}(\vec{R},r)$ that contains {\it asymmetric} bilinear tensor terms proportional to $J_{n, \mu \nu}(\vec{R}) \, J_{n, \nu \mu}(\vec{R})$ solely comes from the tensor interaction~\cite{Bender:2009ty}. This can be easily traced back to the spin-space coupling that characterizes the tensor operator.
\begin{figure}[hptb]
{\includegraphics[keepaspectratio,angle = 0,width=\columnwidth]%
{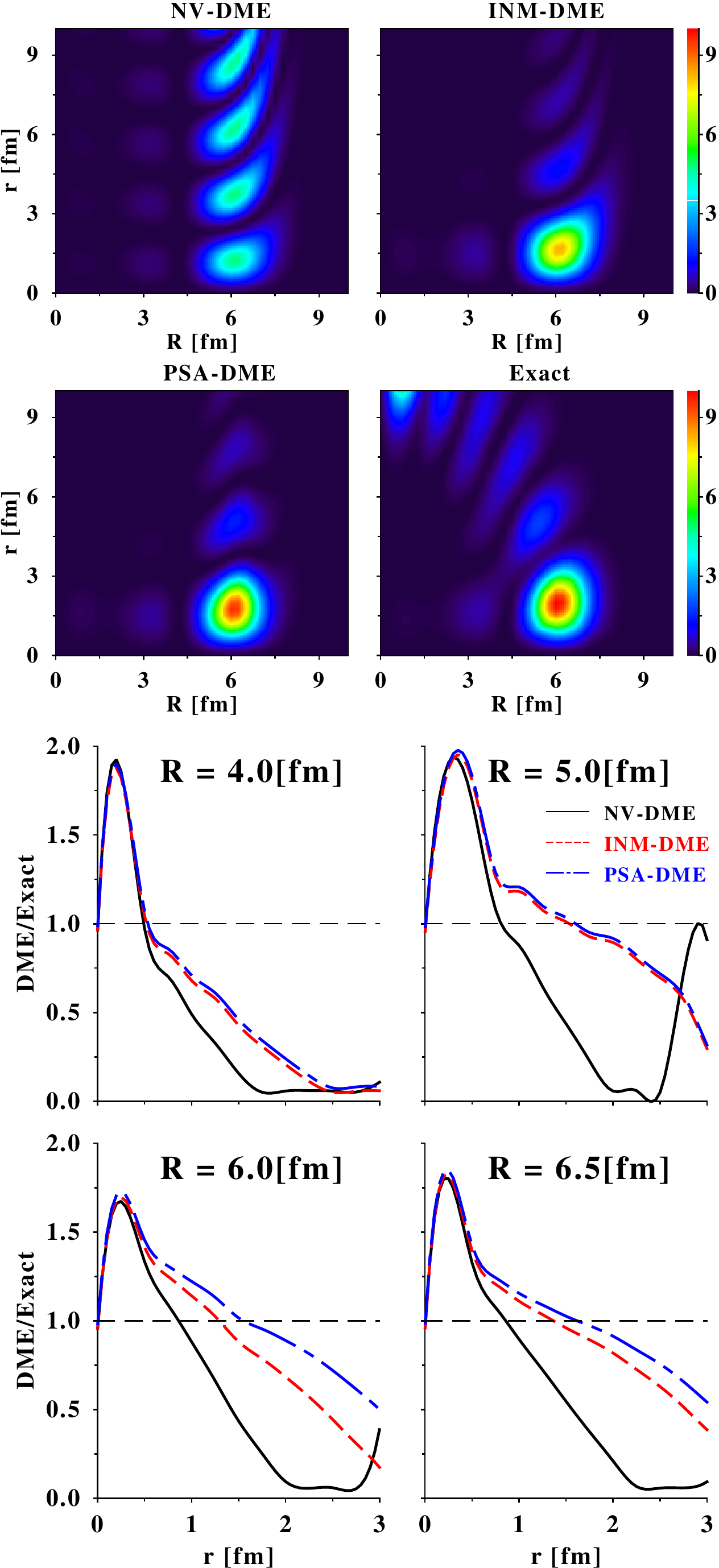}}\caption{ \label{fig:tensor-profile}(Color online) Comparison of $T^{F}_{nn,1}(\vec{R},r)$
and $T^{DME}_{nn,1}(\vec{R},r)$ where the latter is computed from NV-DME, PSA-DME or from PSA-DME with $P^n_2(\vec{R}) =0$ which we denote as INM-DME.   Upper panels: two-dimensional integrands. Lower panels: ratios of $T^{DME}_{nn,1}(\vec{R},r)$ over $T^{F}_{nn,1}(\vec{R},r)$ for fixed values of $R$. Densities are obtained from a converged self-consistent calculation of
$^{208}$Pb with the SLy4 Skyrme EDF in the particle-hole channel and no pairing.}
\end{figure}
Since the numerical tests are presently carried out for spherical systems, we are only concerned with $T^{F}_{nn,1}(\vec{R},r) $ and $T^{DME}_{nn,1}(\vec{R},r)$. For spin-unsaturated nuclei, $T^{F}_{nn,1} (\vec{R},r)$ is highly localized around the nuclear surface as seen in Fig.~\ref{fig:tensor-profile} for $^{208}$Pb. The same figure shows the progressive and significant improvement that the PSA approach brings to the DME of the vector part of the OBDM. Within the typical range of nuclear-interactions, NV-DME falls off much faster than PSA-DME. Less importantly, NV-DME also introduces artificial and pronounced structures in a region that corresponds to the tail of the interaction. Both of these drawbacks are rectified progressively by PSA-DME. While most of the improvement is already brought by the spherical PSA ($P_2(\vec{R})=0$), an even better accuracy is obtained by incorporating the quadrupolar deformation $P_2(\vec{R})$ of the local momentum Fermi distribution. The overestimation of $T^{F}_{nn,1}(\vec{R},r)$ at very small $r$ seen for all DMEs in the lower panels of Fig.~\ref{fig:tensor-profile} corresponds to a region where the integrand is small and where its weight is further reduced in the integrated energy by the $r^2$ phase-space factor.

Coming to the energy itself, i.e. to the integrated product of the interaction $v_T(r)$ with the tensor energy density, we compare
\begin{eqnarray}
E^F_T[nn] &=& 4 \pi \!\int\! d \vec{R} \,d r\, r^2 \, v_T(r)\, T^F_{nn}(\vec{R},r) \, ,
\label{eqn:VCE} \\
E^{DME}_T[nn] &=& 4 \pi \! \int\! d \vec{R} \,d r\, r^2 \, v_T(r)\, T^{DME}_{nn}(\vec{R},r) \, .
\label{eqn:VCD}
\end{eqnarray}
which for spherical nuclei reduce to the contribution from $T^F_{nn,1}$ and $T^{DME}_{nn,1}$.
\begin{figure}[hptb]
{\includegraphics[keepaspectratio,angle = 0,width=0.9\columnwidth]%
{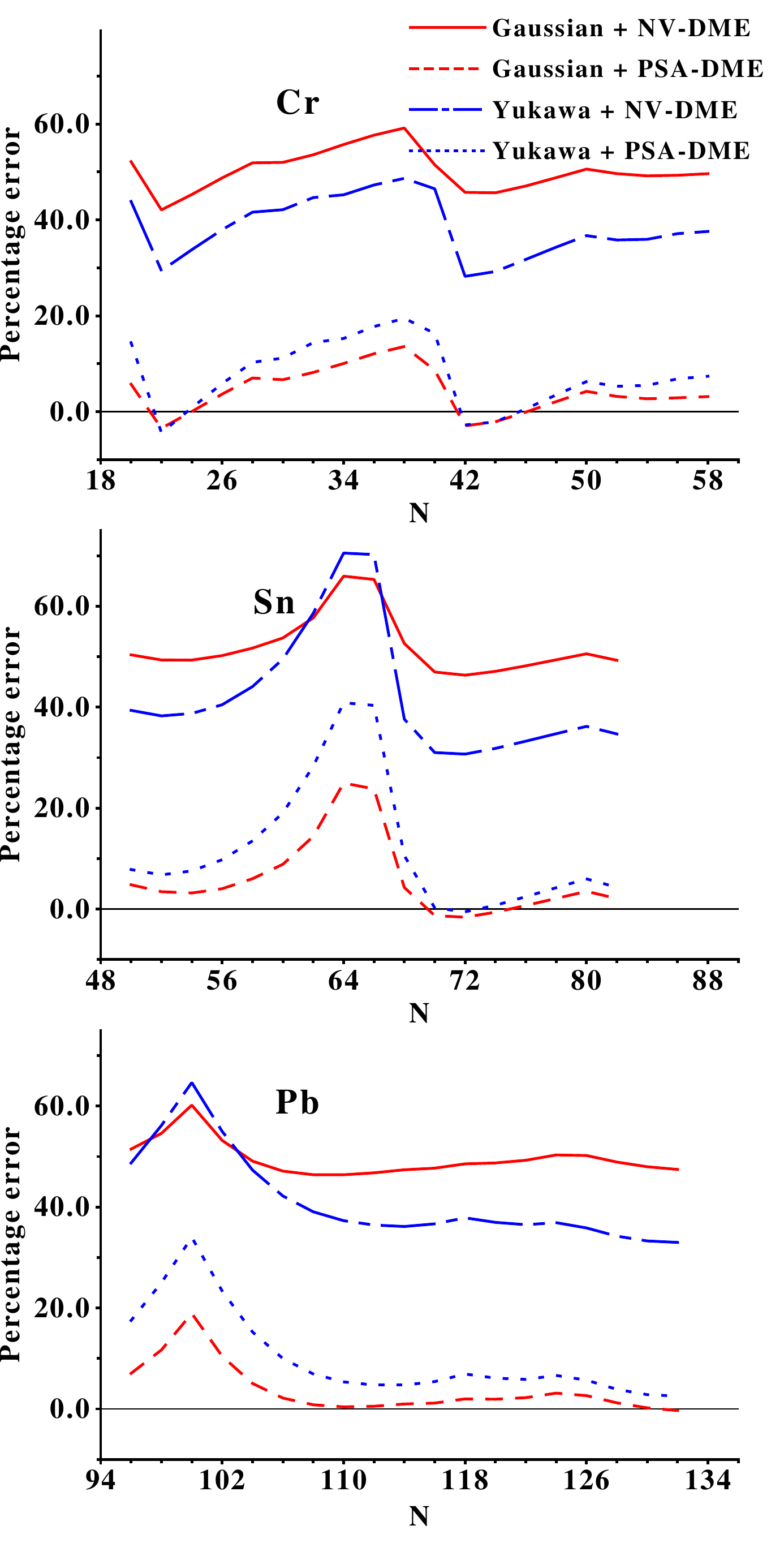}}\caption{ \label{fig:tensor-integrated}(Color online) Percentage error of $E^{DME}_T [nn]$ compared to $E^F_T [nn]$ where the former is either computed from NV-DME or from PSA-DME. Densities are obtained from self-consistent EDF calculations using the SLy4 Skyrme EDF in the particle-hole channel and no pairing. Notice the different vertical scale compared to Fig.~\ref{fig:central-integrated}.}
\end{figure}
Figure~\ref{fig:tensor-integrated} shows the relative error of NV-DME and PSA-DME compared to the exact Fock contribution, for both the Gaussian and the renormalized-Yukawa radial form factors and for three semi-magic isotopic chains. For both types of interaction, the percentage error of NV-DME easily reaches 40\%. This is in contrast to PSA-DME whose percentage error is typically within $\pm10\%$ for most parts of the three isotopic chains. This can be traced to the fact that, while both NV-DME and PSA-DME overestimate the reference quantity for small $r$ (typically less than $1$ fm), NV-DME decreases much faster with $r$, thereby overcompensating for its initial overestimation. In contrast, PSA-DME stays close to the exact value for a much larger range of $r$ values.

\begin{figure}[hptb]
{\includegraphics[keepaspectratio,angle = 0,width=\columnwidth]%
{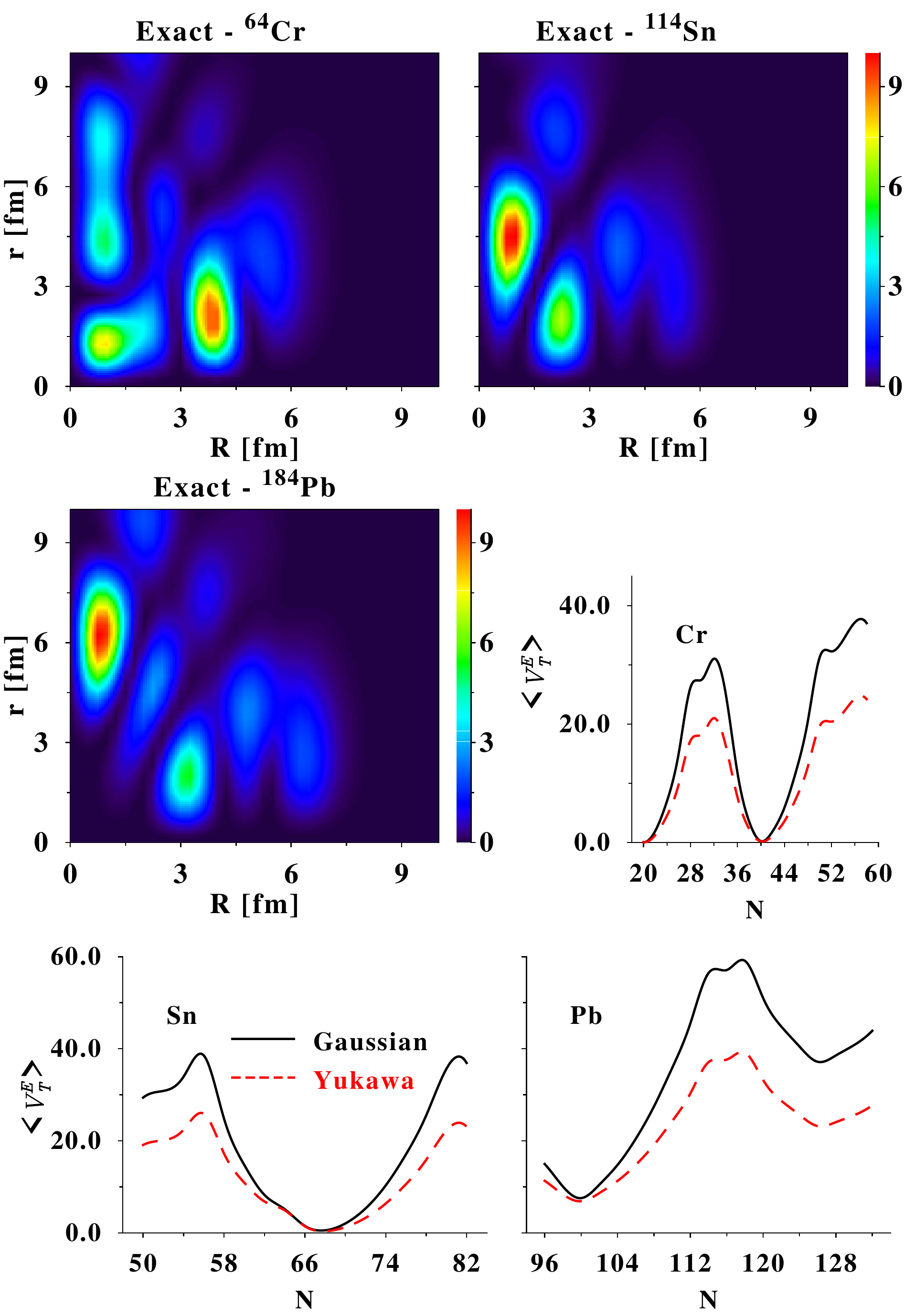}}\caption{ \label{fig:tensor-profile2}
(Color online) A few representative nuclei with diffuse $T^F_{nn,1}(\vec{R},r)$ together with absolute $E^F_T[nn]$ for the corresponding isotopic chains.
Densities are obtained from a self-consistent EDF calculation using the SLy4 Skyrme functional in the particle-hole part and no pairing.}
\end{figure}

There exist short sequences of isotopes for which the percentage error shows a considerable increase. The fact that both DMEs display such a feature suggests that the problem is independent of the specific form of the $\Pi^\vec{s}_1$ function used. To identify the source of the problem, Fig.~\ref{fig:tensor-profile2} shows $T^{F}_{nn,1}(\vec{R},r)$ for three nuclei displaying a sudden loss of accuracy. One notices that $T^F_{nn,1}(\vec{R},r)$ extends over larger intervals in $R$ and $r$ than for $^{208}$Pb (see Fig.~\ref{fig:tensor-profile}). This corresponds to the fact that the selected nuclei are nearly spin-saturated and generates very small $E^F_T[nn]$ in absolute value, as seen from the lower panels of Fig.~\ref{fig:tensor-profile2}. As a result, the relative inaccuracy of any DME becomes large and the percentage error increases suddenly. Of course, the resulting error in the total EDF remains very small as the corresponding tensor contribution is anyway negligible, i.e. the local spin-orbit density $\vec{J}_q(\vec{R})$ is close to zero in nearly spin-saturated nuclei. Eventually, those sudden losses of relative accuracy are not as worrying as Fig.~\ref{fig:tensor-integrated} initially suggests.

In conclusion, the use of PSA techniques has allowed us to bring the DME applicable to the bilinear product of non-local spin densities on the same level of accuracy as for terms depending on the scalar part of the OBDM. One could certainly work even harder to bring the overall DME accuracy below 1\%. This could be achieved (i) by allowing free parameters in the $\Pi-$functions to be optimized on a set of reference calculations\footnote{As will be shown in a future publication, parameterizing $\Pi^{\vec{s}}_1$ cannot remove the sudden loss of relative accuracy discussed above for spin-saturated nuclei. As already stated, this is not a problem in the end as the corresponding contribution to the energy is negligible anyway.} and/or (ii) by going to higher orders in the DME, consistently for both the scalar and the vector parts of the OBDM. This should however be done within the frame of the generalized Skyrme EDF proposed in Ref.~\cite{carlsson09}.

\subsection{Fock contribution from $V_{LS}$}
\label{section:spin-orbit-contribution}

\subsubsection{Basic analysis}

We now turn to the spin-orbit contribution to the Fock energy. As shown in Eq.~\eqref{spin-orbit-contribution}, and unlike for central and tensor forces, such a contribution involves both the scalar and the vector parts of the OBDM. In this case, we first compare the spin-orbit energy density
\begin{eqnarray}
{LS}^F_{nn}(\vec{R},r) &=& \frac{i}{4 \pi} \! \int \! d \vec{e}_r \, \vec{s}_n(\vec{r}_1, \vec{r}_2)
\cdot \vec{r} \times \vec{\nabla}_2 \rho_n (\vec{r}_2, \vec{r}_1) \, \,\,, \label{eqn:LSEprofile}
\end{eqnarray}
to its DME counterpart
\begin{eqnarray}
{LS}^{DME}_{nn}(\vec{R},r) &=&  \frac{1}{6}\,\Pi^{\vec{s}}_{1} (k^n_F r)
\, r^2 \sum^{z}_{\mu,\nu,\beta=x}  \epsilon^{\mu \nu \beta} J_{n,\mu \nu}(\vec{R}) \nonumber\\
&& \times \nabla^{\beta}_\vec{R} \biggl(\Pi^{\rho}_0 (k^n_F r) \rho_n (\vec{R}) \biggr)\,, \nonumber
\end{eqnarray}
which eventually depends on which variants of the DME is being adopted\footnote{The numerical tests shown in the present section actually use INM-DME rather than PSA-DME, i.e. $k^q_F$ is employed rather than $\tilde{k}^q_F$ in $\Pi^{\vec{s}}_{1}$. We still label the results as PSA-DME as no significant difference is seen compared to INM-DME.} and that reduces for spherical systems to
\begin{eqnarray}
{LS}^{DME}_{nn}(\vec{R},r) &=&  \frac{1}{6}\,\Pi^{\vec{s}}_{1} (k^n_F r)
\, r^2\nonumber\\
&& \times \vec{J}_n(\vec{R})\cdot \vec{\nabla}_{\vec{R}} \biggl(\Pi^{\rho}_0 (k^n_F r) \rho_n (\vec{R}) \biggr)\,\,\,. \label{eqn:LSDprofile}
\end{eqnarray}
Note that terms containing more than two
gradients have been truncated in ${LS}^{DME}_{nn}(\vec{R},r)$.

\begin{figure}[hptb]
{\includegraphics[keepaspectratio,angle = 0,width=\columnwidth]%
{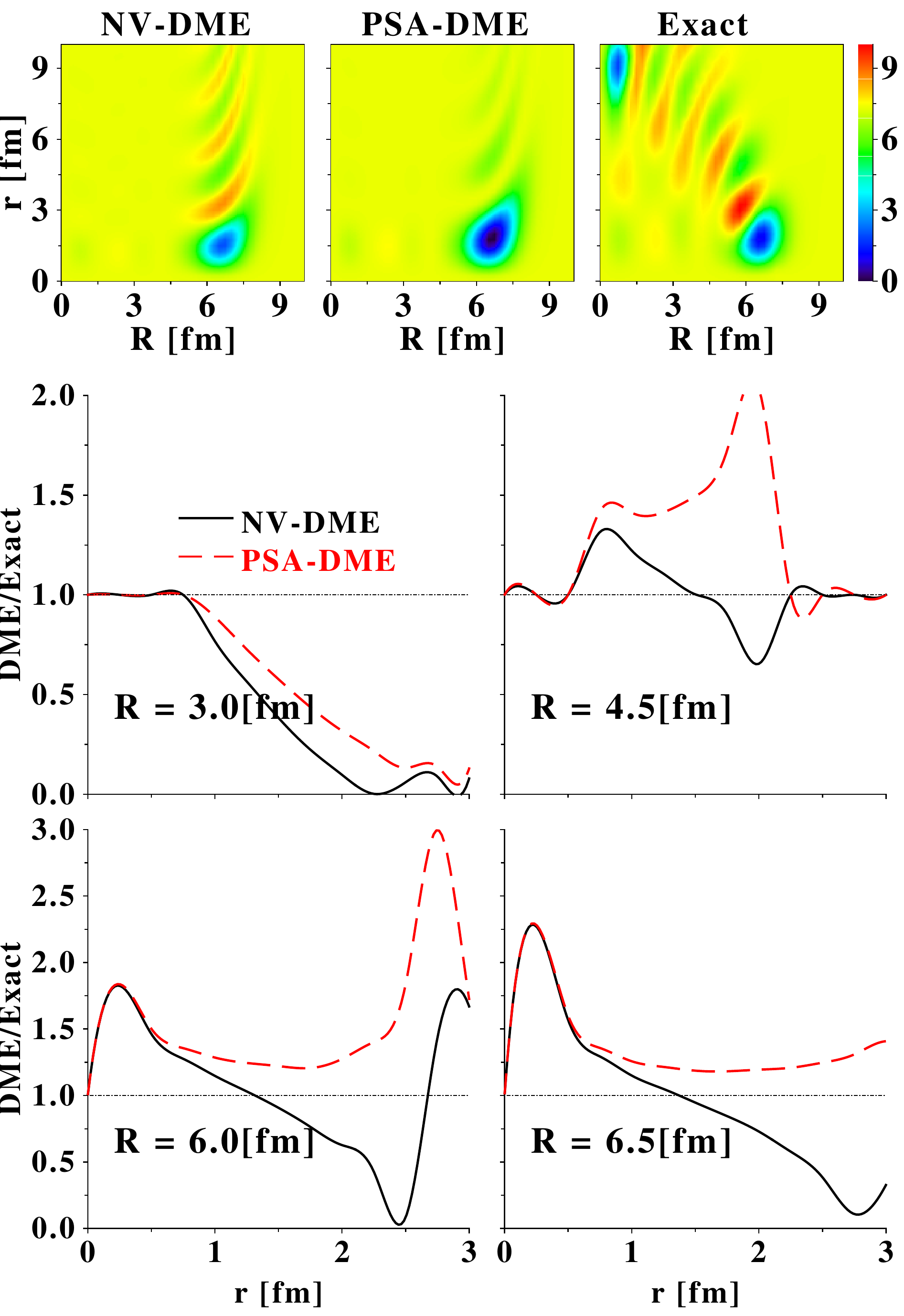}}\caption{ \label{fig:spinorbit-profile}(Color online) Comparison of ${LS}^{F}_{nn}(\vec{R},r)$
and ${LS}^{DME}_{nn}(\vec{R},r)$ where the latter is computed from either NV-DME or PSA-DME. Upper panels: two-dimensional integrands. Lower panels: ratios of ${LS}^{DME}_{nn}(\vec{R},r)$ over ${LS}^{F}_{nn}(\vec{R},r)$ for fixed values of $R$. Densities are obtained from a converged self-consistent calculation of $^{208}$Pb with the SLy4 Skyrme EDF in the particle-hole channel and no pairing.}
\end{figure}

Figure~\ref{fig:spinorbit-profile} shows that PSA-DME significantly overestimates (in absolute values) the maximum peak of ${LS}^{F}_{nn}(\vec{R},r)$ at the nuclear surface. In addition, oscillations at larger $r$, i.e. in the tail of the two-nucleon interaction, are not captured by PSA-DME. In contrast, NV-DME reproduces relatively well the density profile ${LS}^{F}_{nn}(\vec{R},r)$, in particular as for the main peak at the nuclear surface. This suggests that the significant improvement for PSA-DME over NV-DME as to reproducing the tensor energy density does not transpose to the spin-orbit energy density.  The previous assertions are supported by tests carried over several isotonic and isotopic chains. Looking for possible improvements, we tested that including truncated higher-order terms associated with the action of $\vec{\nabla}_\vec{R}$ on  $(1/4 \Delta \rho_n -\tau_n + 3/5 k^{n\,2}_F \rho_n)$, when going from Eq.~\ref{eqn:LSEprofile} to \ref{eqn:LSDprofile}, does not improve the accuracy of PSA-DME.

Coming to the energy itself, i.e. to the integrated product of the interaction $v_{LS}(r)$ with the spin-orbit energy density, we compare
\begin{eqnarray}
E^F_{LS}[nn] &=& 4 \pi \int d \vec{R} \,d r\, r^2 \, v_{LS}(r)\, {LS}^{F}_{nn}(\vec{R},r)
\label{eqn:VCE} \,\,\, , \\
E^{DME}_{LS}[nn] &=& 4 \pi \int d \vec{R} \,d r\, r^2 \, v_{LS}(r)\, r^2 \, {LS}^{DME}_{nn}(\vec{R},r)
\label{eqn:VCD}\,\,\, .
\end{eqnarray}
\begin{figure}[hptb]
{\includegraphics[keepaspectratio,angle = 0,width=0.9\columnwidth]%
{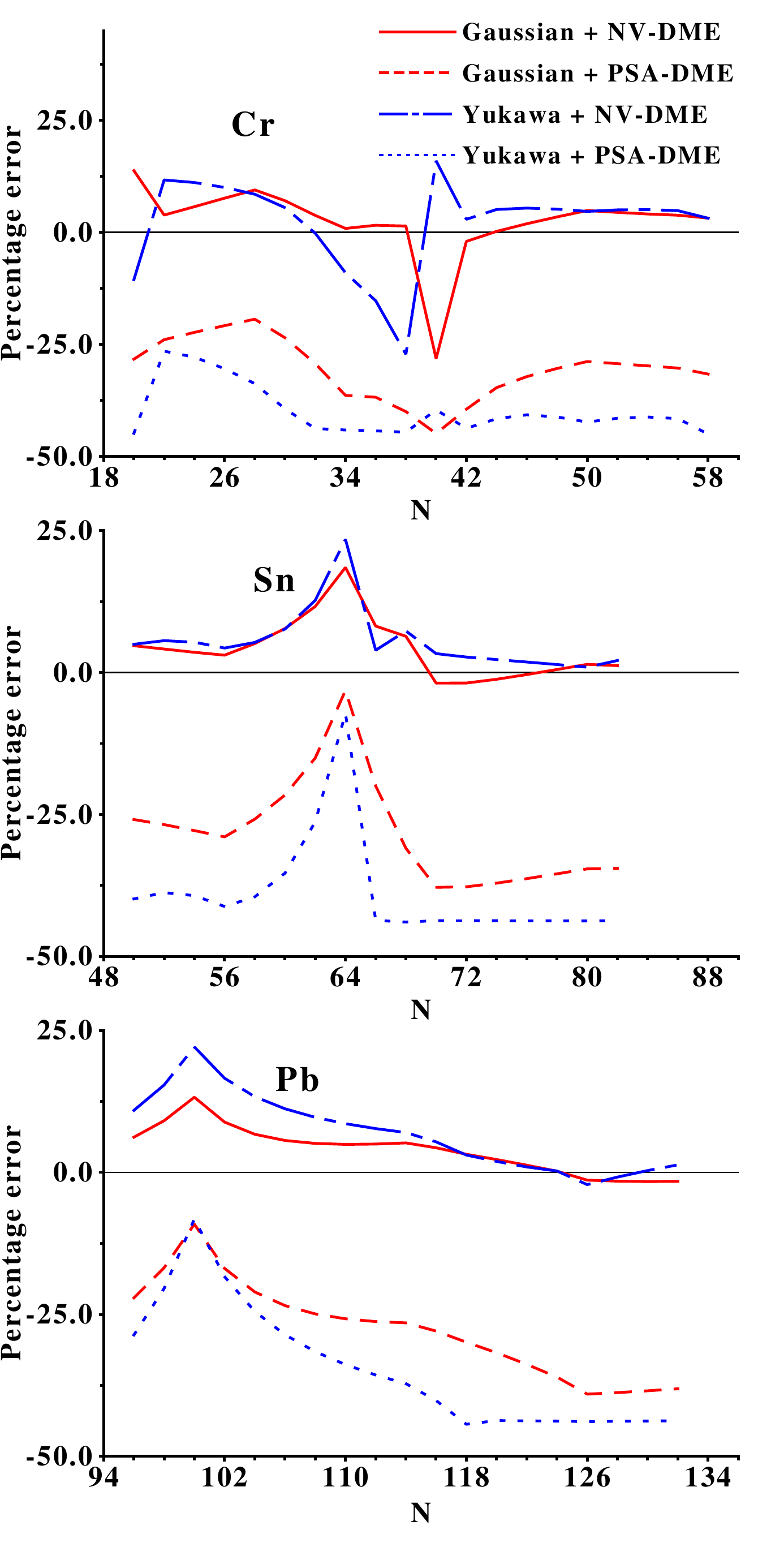}}\caption{ \label{fig:spinorbit-integrated}(Color online) Percentage error of $E^{DME}_{LS} [nn]$ compared to $E^F_{LS} [nn]$ where the latter is either computed from NV-DME or from PSA-DME. Densities are obtained from self-consistent EDF calculations using the SLy4 Skyrme EDF in the particle-hole channel and no pairing. Notice the different vertical scale compared to Figs.~\ref{fig:central-integrated} and~\ref{fig:tensor-integrated}.}
\end{figure}
Figure~\ref{fig:spinorbit-integrated} shows the percentage error obtained for three isotopic chains. In agreement with the analysis done for the  spin-orbit energy density, the percentage error of PSA-DME is impractically large and negative, in the range of -15\% to -50\% for the two schematic interactions used. In contrast, NV-DME provides a much better accuracy with percentage errors within $\pm$ 10\% for most studied isotopes. Last but not least, one notes that the spikes in the percentage errors already discussed in section ~\ref{section:tensor-contribution} arise for the same isotopes and relate to the vanishing non-local spin density in near spin-saturated nuclei.

\subsubsection{Further investigation of the spin-orbit exchange}

The results of the previous section show that NV-DME is better suited than PSA-DME to reproduce the spin-orbit contribution to the Fock energy. This can be confounding in light of the better accuracy obtained using PSA-DME to reproduce the tensor contribution to the Fock energy. We can infer from Fig.~\ref{fig:tensor-profile} that NV-DME underestimates the main peak of the nonlocal spin density while the latter is well captured by PSA-DME. It is thus puzzling to find the opposite for the Fock spin-orbit energy density. In the following we employ a toy model of the OBDM of finite nuclei to show that this is due to a fortuitous cancelation of errors.

Having already a handle on the non-local spin density $\vec{s}_q(\vec{r}_1, \vec{r}_2)$, we focus on the term it multiplies in the spin-orbit energy density, i.e. $\vec{r} \times \vec{\nabla}_2 \rho_q (\vec{r}_1, \vec{r}_2)$, which we first approximate by
$\vec{r} \times \vec{\nabla}_\vec{R} \rho_q (\vec{r}_1, \vec{r}_2)$ thanks to the weak dependence of the non-local matter density on the orientation of $\vec{r}$ ~\cite{martorell84}. Hence, and focusing arbitrarily on neutrons, we want to compare the two quantities
\begin{eqnarray}
G_E &=& \vec{\nabla}_\vec{R} \rho_n(\vec{R},\vec{r})  \, \, \, , \\
G_{DME} &=& \vec{\nabla}_\vec{R} \biggl(\Pi^{\rho}_0(k^n_F r) \, \rho_n(\vec{R})\biggr) \, \, \, ,
\end{eqnarray}
where the latter is independent of whether NV-DME or PSA-DME is used. To do so, we employ a toy model in which the nonlocal and local matter densities are built from a three-dimensional harmonic oscillator model with smeared occupancy~\cite{bhaduri78}. The corresponding analytical expressions, as given in Ref.~\cite{bhaduri78}, read as
\begin{widetext}
\begin{eqnarray}
\rho_n (\vec{R} + \frac{\vec{r}}{2}, \vec{R} - \frac{\vec{r}}{2})  &=& \text{ exp}\biggl[-1/4 \alpha^2 r^2 \frac{1+t}{1-t}\biggr]\,\rho_n(\vec{R})
\label{ExactToyModelNonlocalRho} \, , \\
\rho_n (\vec{R})  & = &\frac{2 \alpha^3}{\pi^{3/2}} (1 -t^2)^{-3/2}\,\text{ exp}\biggl[- \alpha^2 R^2 \frac{1-t}{1+t}\biggr]
\label{ExactToyModelLocalRho}\,,
\end{eqnarray}
where $\alpha^2 \equiv m \omega/\hbar$, and from $\int \rho_n(\vec{R}) \, d \vec{R} = N$, we have
$t \equiv 1-(2/N)^{1/3}$. From Eqs.~\ref{ExactToyModelNonlocalRho} and~\ref{ExactToyModelLocalRho}, one easily obtains
\begin{eqnarray}
\vec{\nabla}_{\vec{R}} \rho_n (\vec{R} + \frac{\vec{r}}{2}, \vec{R} - \frac{\vec{r}}{2}) &=&\,  \text{exp}\biggl[-1/4 \alpha^2 r^2 \frac{1+t}{1-t}\biggr]\,\biggl[\vec{\nabla}_R \rho_n(\vec{R})\biggr]\,\, , \\
\vec{\nabla}_{\vec{R}} \rho_n(\vec{R}) &=& -\frac{4 \alpha^5}{\pi^{3/2}}  (1 -t^2)^{-3/2}\frac{1-t}{1 + t} R \, \text{ exp}\biggl[- \alpha^2 R^2 \frac{1-t}{1+t}\biggr]\,\,\, .
\end{eqnarray}
The corresponding PSA-DME reads
\begin{equation}
\rho_n (\vec{R} + \frac{\vec{r}}{2}, \vec{R} - \frac{\vec{r}}{2})  \approx 3\,\frac{j_1(k^n_F r)}{k^n_F r} \biggl[ 1 +
\frac{r^2}{4} \biggl( -\frac{1+t}{1-t}\,\alpha^2 + \frac{2}{5} k^{n\,2}_F \biggr)\,\biggr] \, \rho_n(\vec{R})\,\label{DMEToyModelNonlocalRho},
\end{equation}
\end{widetext}
such that, given the definition of $k^q_F (\vec{R})$, one can easily obtain
\begin{equation}
\vec{\nabla}_\vec{R} \biggl[\Pi^{\rho}_0(k^n_F r) \rho(\vec{R}) \,\biggr] = j_0 (k^n_F r) \vec{\nabla}_\vec{R}\rho_n (\vec{R})
\end{equation}
and show that
\begin{eqnarray}
G_{ratio}(\vec{R}, \vec{r}) &\equiv&  \frac{G_{DME}(\vec{R}, \vec{r})}{G_E (\vec{R}, \vec{r}) } = j_0(k^n_F r)\, \text{exp}\biggl[1/4  \alpha^2 r^2 \frac{1+t}{1-t}\biggr]\,.\nonumber
\end{eqnarray}
In order to study $G_{ratio}$ quantitatively, we fix the inverse oscillator length, $\alpha$, using the
Blomqvist and Molinari formula~\cite{blomqvist68}, i.e. $1/\alpha^2 = \bigl(0.90\, A^{1/3} + 0.70 \bigr)$.  In subsequent discussions, we take reasonable combinations of $A$ and $N$ although we show that the conclusions of the present section are independent of the actual value of $A$.

\begin{figure}[hptb]
{\includegraphics[keepaspectratio,angle = 0,width=\columnwidth]%
{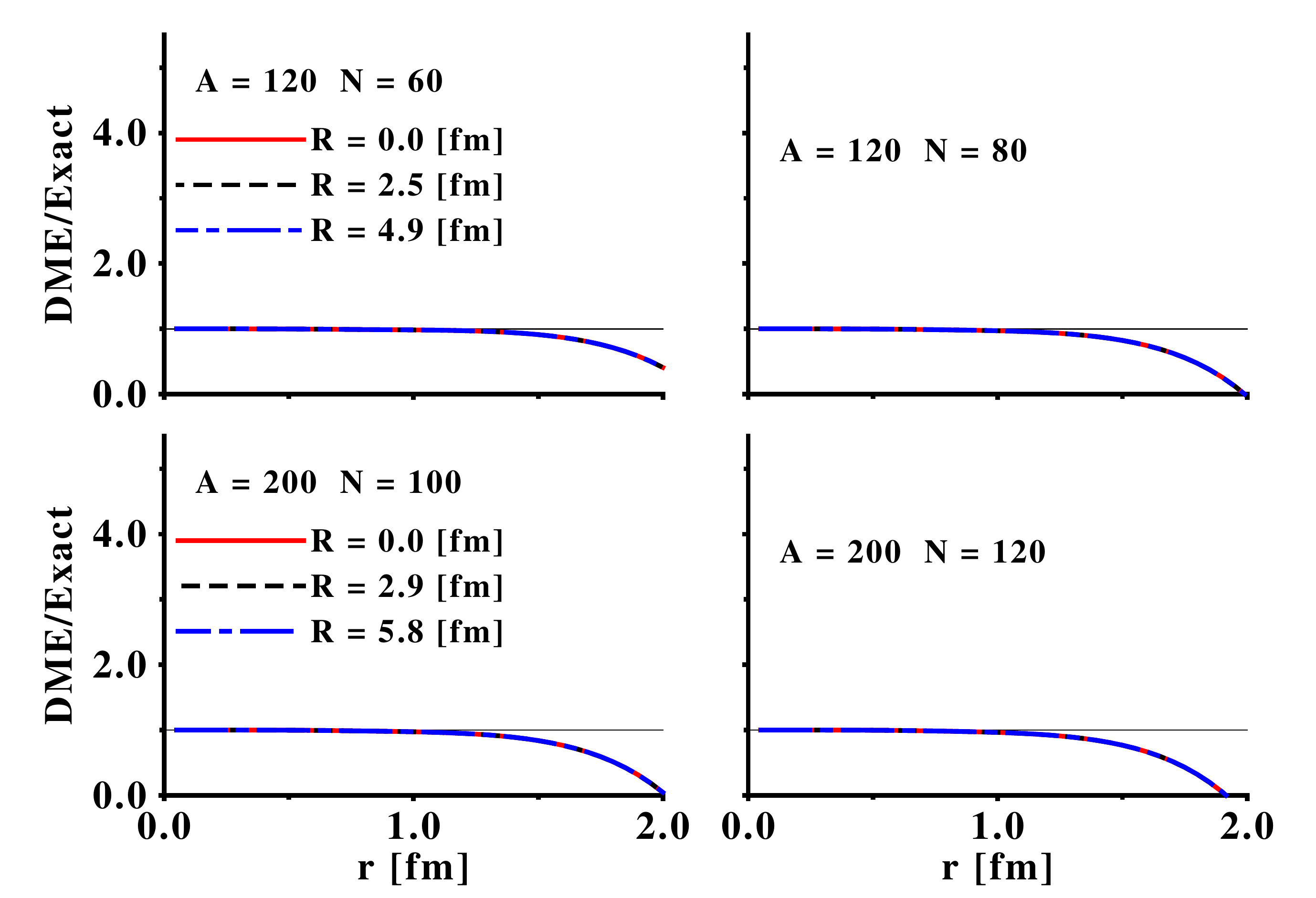}}\caption{\label{fig:WhySpinOrbit691215nograd}(Color online) Ratio of the DME (Eq.\eqref{DMEToyModelNonlocalRho}) over the exact (Eq.\eqref{ExactToyModelNonlocalRho}) expressions of the toy nonlocal matter density.}
\end{figure}

Before analyzing the behavior of $G_{ratio}(\vec{R}, \vec{r})$, it is worth noticing that the toy nonlocal matter density is exactly separable in relative and center-of-mass coordinates. Such a separability being one inherent, usually only approximate, aspect of the DME, we expect the latter to work well in the present case~\cite{bhaduri78}. Computing the same ratio as in $G_{ratio}(\vec{R}, \vec{r})$ {\it without} the gradient operators, we do indeed obtain the good performance of the DME as is visible in Fig.~\ref{fig:WhySpinOrbit691215nograd}. Note in particular that the ratio is independent of the value of $R$. Such a result proves that the toy model provides a situation comparable to the one studied in Sec.~\ref{section:results-central}, i.e. the DME of the scalar part of the density matrix performs well. Such a performance sets the stage in view of qualifying the results obtained below for $G_{ratio}(\vec{R}, \vec{r})$.

\begin{figure}[hptb]
{\includegraphics[keepaspectratio,angle = 0,width=\columnwidth]%
{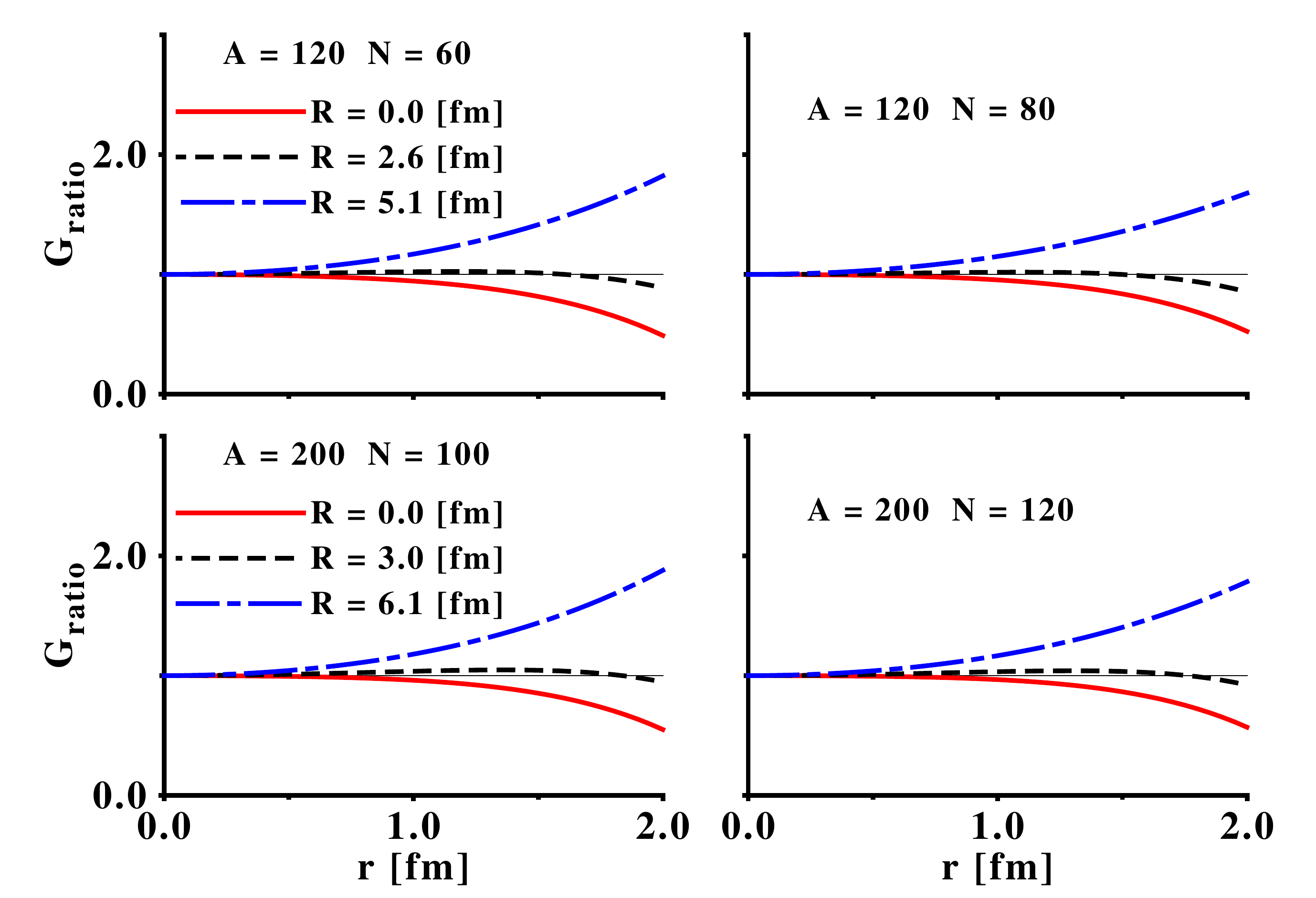}}\caption{\label{fig:Whyso691215}(Color online) $G_{ratio}(\vec{R}, \vec{r})$ as a function of $r$ for a selected set of $(\vec{R},\,A,\,N)$.}
\end{figure}

In order to identify the short distance behavior of $G_{ratio}(\vec{R}, \vec{r})$, we perform a
Taylor series expansion in $r$
\begin{equation}\label{Gerrorseries}
G_{ratio} (\vec{R}, \vec{r}) \approx 1 \,+\, \biggl(-\frac{k_F^{n\, 2}}{6} + \frac{\alpha^2 (1+t)}{4 (1-t)}\biggr) r^2\,.
\end{equation}
Looking close to the surface of
the nucleus, one can neglect $k^{n\,2}_F/6$ in comparison with the second term of Eq.~\eqref{Gerrorseries}.
Defining $G_{error}(\vec{R} , \vec{r}) \equiv G_{ratio}(\vec{R}, \vec{r})-1$, one obtains
\begin{equation}
G_{error} (\vec{R}, \vec{r}) \approx  \frac{\alpha^2 (1+t)}{4 (1-t)}\, r^2 \, .
\label{errorapproxinside}
\end{equation}
Eq.~\eqref{errorapproxinside} is valid around the nuclear surface.
Inside the nucleus, one cannot neglect the first term ($k^{n\,2}_F/6$) of Eq.~\eqref{Gerrorseries}.
This is irrelevant as the spin-orbit energy density is concentrated around the nuclear surface.
Figure~\ref{fig:Whyso691215} bears our expectation i.e. overestimation of $G_E$
 by $G_{DME}$ around the nuclear surface for a wide range of $R$, $A$ and $N$ values. It can also be seen that there
 is a gradual and systematic shift from slight underestimation to
overestimation as one moves from inside the nucleus to the nuclear surface.

Keeping the results shown in Fig.~\ref{fig:WhySpinOrbit691215nograd} as a reference, we conclude that the application of the gradient operator on the scalar part of the density matrix deteriorates the quality of the DME that overestimates the exact results, in particular as one goes to the surface of the nucleus where the exchange spin-orbit energy density is maximum. Combined with the good approximation of the vector part of the density matrix, such a semi-quantitative analysis explains the overall overestimation (in absolute value) of the exchange spin-orbit energy provided by PSA-DME (see Fig.~\ref{fig:spinorbit-integrated}). Contrarily, the underestimation of the vector part of the density matrix by NV-DME provides a fortuitous, but rather accurate, cancelation of errors such that the nonlocal spin-orbit energy density is much better reproduced overall (see Fig.~\ref{fig:spinorbit-integrated}). Even though we can be satisfied with such a situation in the short term future and advocate the use of the NV-DME variant for the spin-orbit contribution to the Fock energy, it would be more satisfying on the long run to design a suitable DME for the gradient of the scalar part of the density matrix that can be combined with the improved PSA-DME for the vector part.

\section{Conclusions and outlook}
\label{conclusions}

The present paper is part of a long-term project to build non-empirical nuclear energy density functionals from realistic two- and three-nucleon interactions using many-body perturbation theory~\cite{Lesinski:2008cd,Drut:2009ce,Duguet:2009gc,Bogner:2008kj}. The density matrix expansion is an important component of this effort, as it can be used to construct numerically-tractable approximations to the non-local Hartree-Fock energy. 
In the first part of this paper, we assessed the accuracy of the DME at reproducing central, tensor, and spin-orbit contributions to the non-local Fock energy. Our central finding is that the conventional DME of Negele and Vautherin performs very poorly in describing the spin-vector part of the density matrix, while the scalar part is described reasonably well. In order to address this deficiency, we have reformulated the density matrix expansion using phase-space averaging techniques. The PSA formulation offers the following benefits:

\begin{itemize}

\item[(i)] It allows one to design expansions of both the scalar and the vector parts of the OBDM on an equal footing. This constitutes a significant improvement over the formulation of Negele and Vautherin who, as they acknowledged in their seminal paper, were not able to provide a satisfactory expansion of the vector part of the density matrix. Considering that the vector part of the density matrix is non-zero in spin-unsaturated nuclei, i.e. in the large majority of nuclei, such an improvement is mandatory in view of constraining a universal energy density functional.

\item[(ii)] By construction, the PSA formulation allows one to incorporate information about the local momentum distribution of the Fermi system of interest. For the scalar part of the OBDM, one recovers the satisfactory expansion of Negele and Vautherin by averaging over the phase space of the locally-equivalent infinite nuclear matter system. For the vector part of the OBDM, one can go beyond this by taking into account the anisotropy that characterizes the local-momentum distribution at the spatial surface of finite Fermi systems. In contrast to the scalar part of the density matrix for which it has little impact, incorporating the deformation of the local momentum distribution in the expansion of its vector part is crucial since the latter peaks at the nuclear surface where such an anisotropy is maximum.
\end{itemize}

In the second part of the paper, we gauged the accuracy of the new PSA-DME and the original NV-DME over a large set of semi-magic nuclei using two non-self consistent measures, i.e., the Fock energy density profile and the Fock energy itself. The different analytical structures of the central, tensor and spin-orbit contributions led us to perform separate tests for each type of contribution. The main conclusions were:
\begin{itemize}
\item[(a)] A few percent accuracy is reached for the central force contribution to the Fock energy that depends on the scalar part of the density matrix. The level of accuracy is insensitive to the particular variant of density matrix expansion.
\item[(b)] For Fock energy contributions from the central and tensor forces that depend on the vector part of the density matrix, the original expansion of Negele and Vautherin leads to about $50\%$ errors. The new expansion based on phase-space averaging techniques reduces errors to the few percent level, which is the same level of accuracy as for terms involving the scalar part of the density matrix only.
\item[(c)] The spin-orbit exchange is somewhat trickier as it combines the vector part of the density matrix with the gradient of its scalar part. Surprisingly, the expansion of Negele and Vautherin is shown to work much better than the new one proposed here. Using a semi-realistic toy model, we demonstrated that this is due to a fortuitous cancelation of errors between the underestimation of the vector part of the density matrix and the overestimation of the gradient of its scalar part. Even though one can be satisfied in the short term with using the NV-DME variant for the spin-orbit contribution to the Fock energy, the present analysis calls for the design of a suitable expansion of the gradient of the scalar part of the density matrix that can be combined with the improved expansion proposed here for the vector part.
\end{itemize}

Optimizing the density matrix expansion for the central, tensor and spin-orbit contributions to the Fock energy as explained above, one reaches an overall error level of a few-percent over a representative set of semi-magic nuclei. With such an accuracy at hand, one can envision using the corresponding generalized Skyrme-like energy functional as a microscopically-constrained starting point around which future refined phenomenological parameterizations can be built. Indeed, the goal of a forthcoming publication~\cite{gebremar09a} is to explicitly compute and analyze all the density-dependent couplings entering the generalized Skyrme-like energy density functional starting from realistic two- and three-nucleon Chiral-EFT potentials at N$^2$LO~\cite{Entem:2003ft,Epelbaum:2004fk}. Of particular interest will be the analysis of (i) the importance of building explicit pion physics into the energy functionals, (ii) the density dependence of spin-orbit and tensor couplings in view of their analysis in recent phenomenological studies~\cite{Lesinski:2007zz,Satula:2008cr,Cao:2009hh,Bender:2009ty} and (iii) the role of three-nucleon forces in these aspects, as well as their effects on the evolution of nuclear shells with isospin. Still, the EDF obtained in this approach will only contain the Hartree-Fock physics such that further correlations must be added to produce any reasonable description of nuclei. In the short term, such an addition can be done empirically by adding the DME couplings to empirical Skyrme functionals and performing a refit of the Skyrme constants to data. While this is a purely empirical procedure, it is motivated by the well-known observation that a Brueckner $G$-matrix differs from the vacuum NN interaction only at short distances. Therefore, one can interpret the refit to data as approximating the short-distance part of the $G$-matrix with a zero-range expansion thru second order in gradients.  Eventually though, it is the goal of a future work to design a generalized DME that is suited to higher orders in perturbation theory.

In addition to using the results of the present and forthcoming papers as building blocks for a microscopically-constrained Skyrme phenomenology, additional work is needed to validate the density matrix expansion method and to gauge its accuracy. Given the outcome of our analysis, several paths can be followed:
\begin{itemize}
\item[(i)] The conclusions reached in the present work must be further validated through self-consistent tests, i.e. binding energies, radii and single-particle energies must be benchmarked against self-consistent Hartree-Fock calculations. The question of whether the Hartree term must be treated exactly is to be addressed quantitatively in such a context.
\item[(ii)] An even better accuracy could be reached for the central and tensor contributions to the Fock energy by going consistently to higher orders in derivatives in the expansion of both the scalar and the vector parts of the density matrix. This should be done within the frame of the extended Skyrme energy density functional proposed in Ref.~\cite{carlsson09}.
\item[(iii)] As already stated, the present analysis of the spin-orbit contribution calls for a suitable expansion of the gradient of the scalar part of the one-body density matrix.
\end{itemize}

%
%

\begin{acknowledgments}
We thank Dick Furnstahl and Jacek Dobaczewski for useful discussions. This work was supported in part by the U.S. Department of Energy UNEDF SciDAC Collaboration under Contract No. DEFC02-07ER41457, and by the U.S. National Science Foundation under Grant Nos. PHY-0456903 and PHY-0758125.
\end{acknowledgments}

%
%

\appendix
\section{Local densities}
\label{appendix:local-densities}

Non-zero local densities can be formed by taking derivatives of the
OBDM up to second order. In the basis from which $| \Phi \rangle$ is built, they read
\begin{eqnarray}
\rho_{q} (\vec{r}) &=&  \sum_{i} \varphi^{\dagger}_{i} (\vec{r} q) \, \varphi_{i} (\vec{r} q) \, \rho^q_{ii}\, , \label{localscalarden} \\
\tau_{q} (\vec{r}) &=&  \sum_{i}  \vec{\nabla}\varphi^{\dagger}_{i} (\vec{r} q) \cdot  \vec{\nabla}\varphi_{i} (\vec{r} q) \, \rho^q_{ii}\, , \label{localkinden} \\
s_{q , \mu}(\vec{r}) & =&  \sum_{ii} \varphi^{\dagger}_{i}
(\vec{r} q ) \, \sigma_{\mu}\,  \varphi_{i} (\vec{r} q) \, \rho^q_{ii}\, , \label{localspinden}\\
j_{q , \mu} (\vec{r}) &=& - \frac{i}{2} \sum_{i} \Bigl( \varphi^{\dagger}_{i}
(\vec{r} q )\, \nabla_{\mu} \, \varphi_{i} (\vec{r} q)\nonumber\\
&&\, -\,\nabla_{\mu} \varphi^{\dagger}_{i} (\vec{r} q ) \, \varphi_{i} (\vec{r} q) \Bigr ) \, \rho^q_{ii}\, , \label{localcurrentden}\\
J_{q, \mu \nu } (\vec{r})&=& -
\frac{i}{2}\,\sum_{i} \,\biggl(\varphi^{\dagger}_{i} (\vec{r}
q)\,\biggl[ \sigma_{\nu} \nabla_{\mu} \varphi_{i} (\vec{r} q)
\biggr] \,\nonumber\\
&& - \biggl[ \nabla_{\mu}\,\varphi^{\dagger}_{i} (\vec{r}
q) \biggr]\,\sigma_{\nu}\,\varphi_{i} (\vec{r} q) \, \biggr) \,\rho^{q}_{ii}\,, \label{localspinorbittensorden}
\end{eqnarray}
\begin{eqnarray}
T_{q , \mu } (\vec{r}) &=& \sum_{i} \,
\vec{\nabla} \varphi^{\dagger}_{i} (\vec{r} q) \,
\biggl[\sigma_{\mu} \, \cdot \vec{\nabla}
\varphi_{i} (\vec{r} q) \biggr]\, \rho^{q}_{ii}\,, \label{localkineticspinden}\\
F_{q, \mu  } (\vec{r}) &=& \frac{1}{2}\,\sum_{i} \,\biggl(\biggl[ \vec{\nabla}\cdot \vec{\sigma}
\varphi^{\dagger}_{i} (\vec{r} q) \biggr] \,\nabla_{\mu} \varphi_{i}
(\vec{r} q) \,\nonumber\\
&& +  \biggl[\nabla_{\mu}\,\varphi^{\dagger}_{i}
(\vec{r} q)\biggr]\, \vec{\nabla}\cdot \vec{\sigma}
\varphi_{i} (\vec{r} q) \, \biggr) \,\rho^{q}_{ii}\, .\label{localtensorkineticden}
\end{eqnarray}
and denote the matter density, the kinetic density, the spin density,
the current density, the spin-current pseudotensor density, the spin kinetic density and
the tensor kinetic density. In the above formulae, $\varphi_{i} (\vec{r} q)$ denotes a spin $1/2$ spinor. Among those local densities, the time-odd ones~\cite{engel75} vanish in time-reversal invariant systems, viz,
\begin{eqnarray}
\vec{s}_q (\vec{r}) &=&0 \, , \quad \quad
\vec{j}_{q} (\vec{r}) =0 \,,  \nonumber \\
\vec{T}_q(\vec{r}) &=& 0 \,,  \quad \quad \vec{F}_q (\vec{r})=0 \, .
\end{eqnarray}

\section{Skyrme-like couplings}\label{appendix:explicit-coupling-forms}

We now provide explicit expressions of the couplings entering the Skyrme-like functional (Eq.\eqref{eqn:skryme-like-EDF})
that results from the application of the DME to the Fock contribution to the ground-state energy
(Eqs. \eqref{central-contribution}, \eqref{spin-orbit-contribution}
and \eqref{tensor-contribution} with the proper coefficients restored). The central, spin-orbit and
tensor parts of the two-nucleon interaction are as specified in section~\ref{section:interaction}.
These couplings are derived under the assumption of time-reversal invariance.
The case where time-reversal invariance is relaxed will be the subject of a future publication.

Starting from the definitions
\begin{eqnarray}
a^{IST}_{1} [ \Pi^{\rho/ \vec{s}
}_i \, \Pi^{\rho/ \vec{s} }_i  ]
&\equiv& 4 \pi  \!\int \!\! d r \, r^2  V^{TS}_I (r)\, \Pi^{\rho/ \vec{s} }_i \, \Pi^{\rho/ \vec{s} }_i  , \label{coupling1}\\
a^{IST}_{2}
[ \Pi^{\rho/ \vec{s} }_i \, \Pi^{\rho/ \vec{s} }_i  ]
&\equiv& \frac{4 \pi}{3} \! \int \!\! d r \, r^4   V^{TS}_I (r) \, \Pi^{\rho/ \vec{s} }_i \, \Pi^{\rho/ \vec{s} }_i , \label{coupling2}
\end{eqnarray}
the couplings take the form
\begin{widetext}
\begin{eqnarray}
{A}^{\rho \rho} &=& + \frac{1}{8}\, a^{C01}_1 \bigl[\Pi^{\rho}_{0}\,\Pi^{\rho}_{0}\bigr] -
\frac{3}{8}\, a^{C11}_1 \bigl[\Pi^{\rho}_{0}\,\Pi^{\rho}_{0}\bigr] \, \nonumber \\
{B}^{\rho \rho} &=& + \frac{3}{16}\,  a^{C10}_1 \bigl[\Pi^{\rho}_{0}\,\Pi^{\rho}_{0}\bigr]+
\frac{1}{16} \, a^{C01}_1 \bigl[\Pi^{\rho}_{0}\,\Pi^{\rho}_{0}\bigr] - \frac{3}{16}\, a^{C11}_1 \bigl[\Pi^{\rho}_{0}\,\Pi^{\rho}_{0}\bigr] + \frac{1}{16} \, a^{C00}_1 \bigl[\Pi^{\rho}_{0}\,\Pi^{\rho}_{0}\bigr] \,  \nonumber \\
{A}^{\rho \tau} &=& - \frac{1}{8}\, a^{C01}_2 \bigl[\Pi^{\rho}_{0} \,\Pi^{\rho}_{2}\bigr] +
\frac{3}{8}\,a^{C11}_2 \bigl[\Pi^{\rho}_{0} \,\Pi^{\rho}_{2}\bigr]  = - 4 \, {A}^{\rho \Delta \rho} \, \nonumber \\
{B}^{\rho \tau} &=& - \frac{3}{16} \, a^{C10}_2\bigl[\Pi^{\rho}_{0} \,\Pi^{\rho}_{2}\bigr] - \frac{1}{16} \, a^{C01}_2\bigl[\Pi^{\rho}_{0} \,\Pi^{\rho}_{2}\bigr] + \frac{3}{16} \, a^{C11}_2\bigl[\Pi^{\rho}_{0} \,\Pi^{\rho}_{2}\bigr] + \frac{1}{16} \, a^{C00}_2 \bigl[\Pi^{\rho}_{0} \,\Pi^{\rho}_{2}\bigr]  = - 4 \, {B}^{\rho \Delta \rho} \, \nonumber \\
{A}^{\rho \nabla J } &=& - \frac{1}{4}\, a^{LS11}_2 \bigl[\Pi^{\rho}_{0} \, \Pi^{\vec{s}}_{1}\bigr] = -  {A}^{\nabla \rho  J } \, \nonumber \\
{B}^{\rho \nabla J} &=& - \frac{1}{8}\, a^{LS10}_2 \bigl[\Pi^{\rho}_{0} \, \Pi^{\vec{s}}_{1}\bigr] + \frac{1}{8}\,
a^{LS11}_2 \bigl[\Pi^{\rho}_{0}
\, \Pi^{\vec{s}}_{1}\bigr]   = - {B}^{\nabla \rho  J} \, \nonumber \\
{A}^{JJ} &=&  - \frac{1}{8} \, a^{C01}_2 \bigl[\Pi^{\vec{s}}_{1}\,\Pi^{\vec{s}}_{1}\bigr] -\frac{1}{8}\,a^{C11}_{2} \bigl[\Pi^{\vec{s}}_{1}\,\Pi^{\vec{s}}_{1}\bigr] +\frac{1}{2}\, a^{T11}_{2} \bigl[\Pi^{\vec{s}}_{1}\,\Pi^{\vec{s}}_{1}\bigr] -
\frac{3}{2}\,a^{T11}_{3} \bigl[\Pi^{\vec{s}}_{1}\,\Pi^{\vec{s}}_{1}\bigr] \, \nonumber\\
{B}^{JJ} &=& + \frac{1}{16} \, a^{C10}_{2} \bigl[\Pi^{\vec{s}}_{1}\,\Pi^{\vec{s}}_{1}\bigr] - \frac{1}{16} \, a^{C01}_{2} \bigl[\Pi^{\vec{s}}_{1}\,\Pi^{\vec{s}}_{1}\bigr] - \frac{1}{16} \, a^{C11}_{2} \bigl[\Pi^{\vec{s}}_{1}\,\Pi^{\vec{s}}_{1}\bigr] + \frac{1}{16} \, a^{C00}_{2} \bigl[\Pi^{\vec{s}}_{1}\,\Pi^{\vec{s}}_{1}\bigr] \, \nonumber \\
 && - \frac{1}{4} \, a^{T10}_{2} \bigl[\Pi^{\vec{s}}_{1}\,\Pi^{\vec{s}}_{1}\bigr]  + \frac{3}{4} \, a^{T10}_{3} \bigl[\Pi^{\vec{s}}_{1}\,\Pi^{\vec{s}}_{1}\bigr] + \frac{1}{4} \, a^{T11}_{2} \bigl[\Pi^{\vec{s}}_{1}\,\Pi^{\vec{s}}_{1}\bigr] - \frac{3}{4}\, a^{T11}_{3} \bigl[\Pi^{\vec{s}}_{1}\,\Pi^{\vec{s}}_{1}\bigr] \, \nonumber\\
A^{J \bar{J}}\,&=&\,- \frac{3}{2}\, a^{T11}_3 \bigl[
\,\Pi^{\vec{s}}_{1}\,\Pi^{\vec{s}}_{1} \, \bigr]\, \nonumber \\
B^{J \bar{J}}\,&=&\, \frac{3}{4}\, a^{T10}_3\, \bigl[
\,\Pi^{\vec{s}}_{1}\,\Pi^{\vec{s}}_{1} \, \bigr]\, -\,
\frac{3}{4}\, a^{T11}_3\,  \bigl[
\,\Pi^{\vec{s}}_{1}\,\Pi^{\vec{s}}_{1} \, \bigr]
\,.\nonumber
\end{eqnarray}
\end{widetext}
To carry on further the computation of the couplings, one must choose an explicit form of the two-nucleon interaction and perform the integrals entering Eqs.~\ref{coupling1} and~\ref{coupling2}. As schematic interactions have been used in the present paper for illustrative purposes, we postpone such an integration to the explicit computation of the couplings obtained from a Chiral-EFT lagrangian at N$^2$LO~\cite{gebremar09a}.

\section{Local anisotropy $P_2(\vec{r})$}
\label{appendix:P_2derivation}

The Husimi distribution is one of the many quantum phase-space
distribution functions. It possesses the key property of positive
definiteness~\cite{husimi40,lee03} and is defined as
\begin{eqnarray}
H_q(\vec{r},\vec{p}) & \equiv & \frac{1}{N} \sum_{i}\biggl{|}\int \!
\varphi_i (\vec{r}_1 q)\,e^{\frac{i}{\hbar}\vec{p} \cdot (\vec{r} - \vec{r}_1)
-\frac{1}{2 r^2_0}(\vec{r} - \vec{r}_1)^2} d \vec{r}_1\biggr{|}^2 \,\nonumber\\
&&\quad\quad\quad\quad \times \, \rho^q_{ii}\,,
\end{eqnarray}
where $N \equiv 1/(\pi^{3/4} r^{3/2}_0)$ and $r_0$ is a chosen parameter. To derive Eq. \eqref{eqn:P_2definition}
for the quadrupolar local anisotropy of the momentum Fermi surface $P^q_2(\vec{r})$
we start from  the definition
\begin{eqnarray}\label{eqn:P_2definition-full}
P^q_2(\vec{r}) &\equiv& \frac{\int d\vec{p}\, \bigl[3(\vec{e}_r \cdot
\vec{p})^2 -\vec{p}^2 \bigr] H_q(\vec{r},\vec{p})}{\int d \vec{p}
\, \vec{p}^2 H_q(\vec{r},\vec{p})}\,,
\end{eqnarray}
and make use of the relations
\begin{eqnarray}
\int d \vec{p} \, \vec{p}^2 \, e^{-\frac{i}{\hbar}\vec{p} \cdot (\vec{r}^{\prime}_1 -\vec{r}_1)} = (2 \pi)^3\hbar^5 \vec{\nabla}^{\prime}_1 \cdot \vec{\nabla} \,
\delta(\vec{r}^{\prime}_1 - \vec{\nabla}_1)\,, \\
e^{-\frac{1}{r^2_0}(\vec{r}_1 -\vec{r}^{\prime}_1)^2} \, \approx \, \delta(\vec{r}_1
-\vec{r}^{\prime}_1) \, + \, \mathcal{O}\bigl((k^q_F r_0)^2\bigr).
\end{eqnarray}
Through direct application of the above relations, one obtains
\begin{eqnarray}
\int d \vec{p} \,\vec{p}^2 \, H_q(\vec{r} , \vec{p})\, &\approx&  \, (2 \pi)^3 \hbar^5\,
\sum_{i}\bigl{|}\vec{\nabla}\varphi_i (\vec{r} q) \bigl{|}^2 \, \rho^q_{ii} \, \nonumber\\
&&+  \mathcal{O}\bigl((k^q_F r_0)^2\bigr)\nonumber \,,\\
\int d \vec{p}\,  \bigl(\hat{r} \cdot \vec{p}\bigr)^2 H_q(\vec{r} , \vec{p})\, &\approx&  \, (2 \pi)^3 \hbar^5\,
\sum_{i} \bigl{|}\bigl(\hat{r} \cdot \vec{\nabla}\bigr)\varphi_i (\vec{r} q) \bigl{|}^2 \, \rho^q_{ii} \nonumber \\
&& + \mathcal{O}\bigl((k^q_F r_0)^2\bigr) \nonumber \,,
\end{eqnarray}
which, plugged into Eq.\eqref{eqn:P_2definition-full}, gives
\begin{eqnarray}\label{eqn:P_2definition-re}
P^q_2(\vec{r}) &=&\biggl[\frac{3}{\tau_q(\vec{r})} \sum_{i} |(\vec{e}_r \cdot \vec{\nabla})
\varphi_i (\vec{r} q)|^2 \,  \rho^q_{ii} -1\biggr]  + \mathcal{O}((k^q_F r_0)^2) \,. \nonumber
\end{eqnarray}

Further simplifications can be performed for spherical systems, using single-particle wave-functions expressed in terms of spherical coordinates $\vec{r}=(r,\theta,\varphi)$ as
\begin{eqnarray}\label{eqn:single-particlewfn}
\varphi_i(\vec{r} q) &=& \frac{u^{q}_{n l j} (r q)}{r}  \sum_{m_l \sigma} Y^{m_l}_{l} (\theta,\varphi)
\langle l m_l \frac{1}{2} \sigma \lvert j m \rangle \, \lvert
\sigma \rangle  ,
\end{eqnarray}
through several angular momentum coupling operations. For that,
the following Clebsch-Gordan and spherical harmonic relations
\begin{eqnarray}
\sum_{\sigma} \, \langle l m_l \frac{1}{2} \sigma \lvert j m \rangle
 ^2  &=&  \frac{2 j + 1 }{2 l + 1 }\,,\\
 \sum_{m_l} \, Y^{m_l \ast}_{l} (\theta',\varphi') Y^{m_l }_{l}
(\theta,\varphi)  &=&  \frac{2 l + 1 }{4 \pi } P_{l}\bigl(\vec{e}_{r'} \cdot \vec{e}_{r}\bigr),\,
\end{eqnarray}
turn out to be handy. In these relations, $Y^{m_l}_l$ refers
to a spherical harmonic function and $P_l$ refers to Legendre polynomial of order l.
Applying these relations, one obtains
\begin{eqnarray}
\sum_{i}\bigl{|}\bigl(\vec{e}_{r} \cdot \vec{\nabla}\bigr)\varphi_i (\vec{r} q)
 \bigl{|}^2 \, \rho^q_{ii} &=&\sum_{n l j } \frac{2 j + 1}{4 \pi}\,
 \biggl(\frac{\partial}{\partial r}\frac{u^{q}_{n l j}(r)}{r} \biggr)^2 \,  \rho^{qnjl}  \nonumber\\
\sum_{i} \bigl{|}\vec{\nabla} \varphi_i (\vec{r} q)
 \bigl{|}^2 \,  \rho^q_{ii} &=&\sum_{n l j }  \frac{2 j + 1}{4 \pi}\,
 \biggl(\frac{\partial}{\partial r}\frac{u^{q}_{n l j}(r)}{r}
\biggr)^2 \,   \rho^{qnjl}  \nonumber\\
&&\, + \sum_{n l j } F(l,j) \,\biggl(\frac{u^{q}_{n l j}(r)}{r^2}
\biggr)^2 \,   \rho^{qnjl}  \nonumber,
\end{eqnarray}
where $F(l,j)$ is some function of $l$ and $j$. The occupation probability of a given spherical shell  $ \rho^{qnjl}$ is one or zero, except for open-shell semi-magic nuclei where the so-called filling approximation provides the valence shell with a partial occupation. To obtain the explicit form of $F(l,j)$, one can use the relation
\begin{eqnarray}
\nabla_{\mu_1} Y^{m^{\prime}}_{l}(\theta,\varphi) \, &=& \,
\frac{1}{r}\,\sum_{LM} \, f(l , L) \, \langle l 1 m^{\prime} \mu_1 |
L M \rangle \, Y^{M}_{L} (\theta,\varphi)\,\nonumber ,
\end{eqnarray}
where
  \[ f(l,L)\,= \,\left\{\begin{array} {ll} - l \,\sqrt{\frac{l +
1 }{2 l + 3}} \,\, \quad \quad\text{ if  $L =l +
      1$} ; \\
- (l + 1 ) \,\sqrt{\frac{l  }{2 l - 1}} \, \,\quad \quad \text{ if
$\,L=l
- 1$} ; \\
0 \, \, \quad \quad \text{otherwise}.  \end{array} \right.
\]
and perform involved angular momentum coupling operations. Alternatively, one notes
that $\sum_{i} \bigl{|}\vec{\nabla} \varphi_i (\vec{r} q)
 \bigl{|}^2 \,  \rho^q_{ii}$ is nothing but the kinetic energy density given in Eq.\eqref{localkinden}
 and use the corresponding expression~\cite{vautherin72a}. Either way, one obtains
\begin{equation}
F(l,j) = \frac{l ( l + 1) (2 j + 1)}{4 \pi}\,.
\end{equation}
Plugging these intermediate results into Eq. \eqref{eqn:P_2definition-re} yields
the expression of $P_2 (\vec{r})$ as
\begin{widetext}
\begin{eqnarray}\label{eqn:P_2spherical}
P_2(\vec{r}) &=& \frac{1}{\tau_q(\vec{r})} \sum_{n l j} \frac{2 j + 1}{4 \pi}\,
\biggl[2 \biggl(\frac{\partial}{\partial r}\frac{u^{q}_{n l j}(r)}{r} \biggr)^2 -
\frac{l (l +1)}{r^2}\biggl(\frac{u^{q}_{n l j}(r)}{r}\biggr)^2 \biggr] \, \rho^{qnjl} \,\,\, ,
\end{eqnarray}
where
\begin{eqnarray}
\tau_q(\vec{r}) &=&\sum_{n l j} \frac{2 j + 1}{4 \pi}\,
\biggl[ \biggl(\frac{\partial}{\partial r}\frac{V^{q}_{n l j}(r)}{r}
\biggr)^2 + \biggl(\frac{V^{q}_{n l j}(r)}{r^2}
\biggr)^2 \biggr] \,  \rho^{qnjl} \,\,\,  .
\end{eqnarray}
\end{widetext}

\section{Scalar part of the OBDM}
\label{appendix:scalar-derivation}

We start from Eq. \eqref{eqn:second-order-expanded}, average over the orientation of $\vec{k}$ and $\vec{r}$\footnote{The order
of the two averaging operations is dictated only by the requirement of simplicity. In this case, we
averaged over the orientation of $\vec{r}$ followed by that of $\vec{k}$.},
and apply relations
\begin{eqnarray}
\frac{1}{4 \pi} \int d \vec{e}_r \, (\vec{r} \cdot \vec{A})
(\vec{r} \cdot \vec{B}) & =&  \frac{ r^2}{3} \vec{A} \cdot \vec{B}\,,\\
\biggl( \nabla^2 + \nabla^{\prime 2 }
\biggr) \rho(\vec{r},\vec{r}^{\,\prime})
\bigg{|}_{\vec{r}=\vec{r}'} &=&\nabla^2 \rho(\vec{r})
-2 \, \tau(\vec{r})\,,
\end{eqnarray}
to obtain
\begin{eqnarray}
\label{taylor}
\rho_{q} \bigl(\vec{R} + \frac{\vec{r}}{2},\vec{R}-\frac{\vec{r}}{2}
 \bigr)& \approx & j_0 (k r) \, \rho_q(\vec{R}) + L(k r) \, \rho_q (\vec{R})\\
 && +  \frac{r^2}{24} \, j_0(k r) \,\bigl( \Delta \rho_q(\vec{R}) - 4 \tau_q (\vec{R}) \bigr)\,, \nonumber
\end{eqnarray}
where
\begin{equation}
L(k r) \, \equiv\,  2  k  r \,j_1 (k r) -\frac{(k r)^2}{2}\,  j_0 (k r)\,.
\end{equation}

As discussed in section~\ref{subsection:scalar-application}, the effects of
anisotropy and diffuseness are minimal for the scalar part of the OBDM.
Therefore, we perform the PSA over the phase-space of the locally equivalent pure-isospin nuclear matter\footnote{The
angle integration with respect to the orientation of $\vec{k}$ is trivial as such a dependence has already been averaged out.} to obtain
\begin{eqnarray}
\label{taylor}
\rho_{q} \bigl(\vec{R} + \frac{\vec{r}}{2},\vec{R}-\frac{\vec{r}}{2}
 \bigr)& \approx & 3 \, \frac{j_1 (k^q_F r)}{k^q_F r} \rho_q(\vec{R}) \,\nonumber\\
 && +
 \frac{r^2}{2} \, \frac{j_1(k^q_F r)}{k^q_F r} \, \varrho_q(\vec{R})\,,
\end{eqnarray}
with the second-order correction density being composed of
\begin{eqnarray}
\varrho_q(\vec{R}) \equiv \frac{1}{4} \Delta \rho_q(\vec{R}) -
  \tau_q (\vec{R}) +  \frac{3}{5} k^{q \, 2}_F \, \Lambda(k^q_F r) \rho_q (\vec{R})\,.
\end{eqnarray}
Expanding $\Lambda (k^q_F r)$ in Taylor series, one has
\begin{equation}
\Lambda (k^q_F r) \approx 1 + \mathcal{O}((k^q_F r)^2)\,.
\end{equation}
such that, by retaining the lowest order only, one recovers Eq.~\ref{DMEscalar} with the $\Pi-$functions given by Eqs.~\ref{pi-functions:scalar0} and~\ref{pi-functions:scalar2}.

\section{Vector part of the OBDM}
\label{appendix:vector-derivation}

We start from Eq. \eqref{eqn:first-order-expandedvector}. For time-reversal invariant systems, the local spin density $\vec{s}_q(\vec{r})$ vanishes. Consequently, the only non-vanishing contribution relates to the term
$\vec{r} \cdot (\vec{\nabla}_1\!-\!\vec{\nabla}_2)$. Using the definition for the
local spin-current pseudotensor density given by Eq. \eqref{localspinorbittensorden},
one obtains
\begin{eqnarray}\label{eqn:basic-vectorPSA-start}
s_{q,\nu} \biggl(\vec{R}+ \frac{\vec{r}}{2} , \vec{R}-
\frac{\vec{r}}{2} \biggr) & \approx & i\, e^{i \vec{r}\cdot \vec{k}}
\, \sum_{\mu} r_{\mu} \, J_{q,\mu \nu} (\vec{R})\,.
\end{eqnarray}
The final step involves performing the PSA over a deformed sphere that characterizes the local momentum distribution.
Let us start from a spheroid given in momentum space given by the equation
\begin{equation}\label{eqn:spheroid}
\frac{k^2_x}{a(\vec{R})^2} + \frac{k^2_y}{a(\vec{R})^2} + \frac{k^2_z}{c(\vec{R})^2}=1\,.
\end{equation}
For ease of notation, we write $a(\vec{R})$ as $a$ and
$c(\vec{R})$ as $c$ in the following. We constrain the position-dependent quantities $a$ and $c$ by
requiring that the spheroid has a given volume and quadrupole moment, viz,
\begin{eqnarray}
V_q &\equiv&\frac{4}{3} \pi^3 k^{q \, 3}_F = \frac{4}{3} \pi^3 a^2 c  \,,\\
 P^q_2(\vec{R})  &=& \frac{2\,(-a^2 + c^2) }{2\, a^2 + c^2 }\, .
\end{eqnarray}
The $\Pi-$function is obtained via the integration over the phase space of interest
\begin{eqnarray}
\Pi^{\vec{s}}_1  = \frac{3}{4 \pi^3 k^{q\,3}_F} \int_{V_q} d \vec{k} \,
e^{i \vec{r} \cdot \vec{k}}\,.
\end{eqnarray}
Carrying out the integration over the volume $V_q$ encompassed by the spheroid given in
Eq. \eqref{eqn:spheroid} can be done by using a stretched coordinate system
from the transformation
\begin{equation}
\vec{k} \equiv (k_x , k_y , k_z) \rightarrow \vec{k}^{\prime}
 \equiv (k_x , k_y , \frac{a}{c} k_z) \,,
\end{equation}
such that one finally obtains
\begin{equation}\label{basicvectorexpansionbis}
\vec{s}_{q,\nu} \biggl(\vec{R}+ \frac{\vec{r}}{2} , \vec{R}-
\frac{\vec{r}}{2}\biggr)  \simeq   i \,
\Pi^{\vec{s}}_1 (k^q_F r) \,\sum^{z}_{\mu=x} r_{\mu} J_{q, \mu \nu} (\vec{R}) \,,
\end{equation}
where
\begin{eqnarray}\label{pis1function_PSA-re}
\Pi^{\vec{s}}_1 (\tilde{k}^q_F r) &\equiv& 3 \, \frac{ j_1 (\tilde{k}^q_F r)}{\tilde{k}^q_F r}\,,
\end{eqnarray}
and
\begin{eqnarray}
\tilde{k}^q_F &\equiv& \biggl(\frac{2 + 2 \,P^q_2(\vec{R})}{2-P^q_2(\vec{R})}\biggr)^{1/3} k^q_F\,.
\end{eqnarray}
Setting $P^q_2(\vec{R})= 0$, which consists of performing the PSA over INM phase-space,
results in the same $\Pi-$function with $\tilde{k}^q_F$ replaced by $k^q_F$.

For spherical systems, one can simplify the expression further
by writing $J_{q,\mu \nu} (\vec{R}) $ as a sum of
pseudoscalar, vector and (antisymmetric) traceless tensor parts
\begin{eqnarray}
J_{q , \mu \nu} (\vec{R}) &=& \frac{1}{3} \, \delta_{\mu
\nu} \, J^{(0)}_{q} (\vec{R}) +  \frac{1}{2} \,
\sum^{z}_{k=x}\epsilon_{\mu \nu k}\, J^{(1)}_{q,k} (\vec{R}) \nonumber\\
&&\quad\quad +  J^{(2)}_{q ,\mu \nu} (\vec{R})\,,
\end{eqnarray}
where the three components read
\begin{eqnarray}
J^{(0)}_{q} (\vec{R}) &\equiv& \sum^{z}_{\mu,\nu=x}\,\delta_{\mu \nu}\,
J_{q , \mu \nu} (\vec{R}) \, ,\\
J^{(1)}_{q,k} (\vec{R}) &\equiv& \sum^{z}_{\mu,\nu=x} \epsilon_{\mu \nu k} \, J_{q ,
\mu \nu} (\vec{R}) \, ,\\
J^{(2)}_{q ,\mu \nu} (\vec{R}) &\equiv& J_{q , \mu \nu} (\vec{R}) - \frac{1}{3} \, \delta_{\mu \nu} \, J^{(0)}_{q} (\vec{R}) \,\nonumber\\
&& - \frac{1}{2} \, \sum^{z}_{k=x}\epsilon_{\mu \nu
k}\, J^{(1)}_{q,k} (\vec{R}) \,.
\end{eqnarray}
In spherical systems, both the pseudoscalar and the tensor parts vanish such that one obtains
\begin{equation}\label{basicvectorexpansion}
\vec{s}_{q} \biggl(\vec{R}+ \frac{\vec{r}}{2} , \vec{R}-
\frac{\vec{r}}{2}\biggr) \, \simeq \,  - \frac{i}{2}\, \Pi^{\vec{s}}_1 (\tilde{k}^q_F r) \, \vec{r} \times \vec{J}_{q} (\vec{R}) \,.
\end{equation}

\bibliography{projdiverg}

\end{document}